\documentclass[aps,prx,twocolumn,showpacs,citeautoscript,superscriptaddress,longbibliography,nofootinbib]{revtex4-1}

 \usepackage{graphicx}  % needed for figures
 \usepackage{epstopdf}
 \usepackage{dcolumn}   % needed for some tables
 \usepackage{bm}        % for math
 \usepackage{amsmath}
 \usepackage{amssymb}   % for math
 \usepackage{wrapfig}
 \usepackage{array}
 \usepackage{comment}
 \usepackage[caption=false]{subfig}
 \usepackage[bookmarks=false,linkcolor=green,urlcolor=blue,colorlinks,citecolor=red]{hyperref}
 \usepackage{exscale,relsize}
 \usepackage[usenames, dvipsnames]{color}
 
\newcommand{\be}{\begin{equation}}
\newcommand{\ee}{\end{equation}}
\newcommand{\beq}{\begin{eqnarray}}
\newcommand{\eeq}{\end{eqnarray}}
\newcommand{\ba}{\[\begin{aligned}}
\newcommand{\ea}{\end{aligned}\]}

\newcommand{\la}{\langle}
\newcommand{\ra}{\rangle}

\newcommand{\V}{{\cal V}}

\renewcommand{\vec}[1]{{\bf #1}}
\renewcommand{\hat}[1]{{\bf {\widehat #1}}}
\renewcommand{\phi}{\varphi}
\renewcommand{\epsilon}{\varepsilon}

\def\nn{\nonumber}
\definecolor{applegreen}{rgb}{0.55, 0.71, 0.0}

\begin{document}

%\title{Giant geometric photoresponse in twisted bilayer graphene: interaction enhanced shift currents}
\title{Shift-current response as a probe of quantum geometry and electron-electron interactions in twisted bilayer graphene}

\author{Swati Chaudhary$^\ddagger$}

\email[]{swatich@caltech.edu}

\affiliation{Department of Physics, California Institute of Technology, Pasadena CA 91125, USA}
\affiliation{Institute for Quantum Information and Matter, California Institute of Technology, Pasadena CA 91125, USA}

\author{Cyprian Lewandowski$^\ddagger$}
\email[]{cyprian@caltech.edu}
\affiliation{Department of Physics, California Institute of Technology, Pasadena CA 91125, USA}
\affiliation{Institute for Quantum Information and Matter, California Institute of Technology, Pasadena CA 91125, USA}

% Maybe
% \author{Stevan Nadj-Perge}
% \affiliation{Institute for Quantum Information and Matter, California Institute of Technology, Pasadena  CA 91125, USA}
% \affiliation{T. J. Watson Laboratory of Applied Physics, California Institute of Technology, 1200 East California Boulevard, Pasadena CA 91125, USA}

\author{Gil Refael}
\affiliation{Department of Physics, California Institute of Technology, Pasadena CA 91125, USA}
\affiliation{Institute for Quantum Information and Matter, California Institute of Technology, Pasadena CA 91125, USA}

%\date{March 2021}

\begin{abstract}
Moir\'e materials, and in particular twisted bilayer graphene (TBG), exhibit a range of fascinating phenomena, that emerge from the interplay of band topology and interactions. We show that the non-linear second-order photoresponse is an appealing probe of this rich interplay. A dominant part of the photoresponse is the shift-current, which is determined by the geometry of the electronic wave-functions and carrier properties, and thus becomes strongly modified by electron-electron interactions. We analyze its dependence on the twist angle and doping, and investigate the role of interactions. In the absence of interactions, the response of the system is dictated  by two energy scales: the mean energy of direct transitions between 
the hole and electron flat bands, and the gap between flat and dispersive bands.  Including electron-electron interactions, both enhance the response at the non-interacting characteristic frequencies as well as produce new resonances. We attribute these changes to the filling-dependent band renormalization in TBG. Our results highlight the connection between non-trivial geometric properties of TBG and its optical response, as well as demonstrate how  optical probes can access the role of interactions in moi\'re materials. 
\end{abstract}

\maketitle
\def\thefootnote{$\ddagger$}\footnotetext{These authors contributed equally to this work.}
\section{Introduction}

Twisted bilayer graphene (TBG) is an exciting arena where quantum geometry and enhanced electronic interactions play both against and with each other. While the interactions are boosted by the flatness of the electronic bands near charge neutrality, geometric effects are amplified by the large size of the Moir\'e unit cell as the lattice constant sets the scale for the Berry connection. This conjunction of interactions and geometry is responsible for a growing list of fascinating effects~\cite{cao2018_1,efetov2019,lu2019superconductors,serlin2019intrinsic,Sharpe605,cao2018_2,yankowitz2019,Balents2020,Zondiner2020,wong2020cascade,serlin2020intrinsic} ranging from surprisingly strong superconductivity~\cite{cao2018_2,yankowitz2019,Balents2020}, to cascade effects~\cite{Zondiner2020,wong2020cascade}, and anomalous Hall phases~\cite{serlin2020intrinsic}. In this work we focus on the second order photoresponse of TBG, and on the shift-current in particular. We contend that it is a unique probe that can wield the enhanced geometric effects of the electronic wave function to systematically probe the role of interactions in TBG at a range of fillings and twist angles. 

By quantum geometry (QG), we refer to the structure of the electronic Bloch wavefunctions. Many interesting signatures of QG are revealed in transport properties and optical responses of these systems~\cite{Xiao2010,morimoto2016topological,Ahn2020,ma2021topology,orenstein2021topology,ahn2021riemannian,topp2021light,osterhoudt2019colossal,chan2016,yang2017divergent,isobe2020high}, and especially in the zero-magnetic field quantized anomalous linear hall effect in setups with time-reversal symmetry (TRS)~\cite{Haldane1988,serlin2020intrinsic,Sharpe605}. The effects of QG go well beyond linear response, and can manifest themselves in nonlinear optical responses (NLOR) as shown recently~\cite{xu2018,wu2017giant,Juan2017,wang2019,moore2010confinement,sodemann2015quantum,young2012,Rangel2017,Tan2016, tan2016shift,cook2017design,wu2017giant,xu2018,liao2021nonlinear,zhang2020giant,ortix2021nonlinear,Pantalean2021,he2021giant}. Furthermore, the NLOR does not require broken TRS, but rather a non-zero Berry curvature profile. These non-linear effects can manifest in various ways, such as non-linear response to DC fields (induced by Berry curvature dipole~\cite{moore2010confinement,sodemann2015quantum,xu2018,liao2021nonlinear,zhang2020giant,ortix2021nonlinear,Pantalean2021,he2021giant}), second-harmonic generation (SHG), and bulk-photovoltaic effects like shift-current (SC)~\cite{young2012,tan2016shift,Ogawa2017,Rangel2017}, and circular photogalvanic effects (CPGE)\cite{Hosur2011,Chan2017,Juan2017}. Recently, there has been a lot of emphasis on the non-linear response to AC fields~\cite{wu2017giant,Chan2017,de2017quantized} which not only serve as a probe of non-trivial topology but also heralds promises of more efficient and robust photovoltaic devices~\cite{cook2017design}.

The shift current response\cite{Baltz1981,belinicher1982kinetic,sipe2000,sturman2020ballistic}, is a particularly interesting part of the NLOR. In topological systems, it could generate a giant DC response from a weak linearly polarized electromagnetic fields, which makes it relevant for photovoltaic applications~\cite{Tan2016, tan2016shift,cook2017design,young2012,Rangel2017}. Furthermore, the shift current response is tied to the quantum geometric properties of the system~\cite{morimoto2016topological, PhysRevX.10.041041,ahn2021riemannian,PhysRevResearch.2.033100} and microscopically arises due to change in properties of the Bloch wavefunction upon excitation between bands. Specifically, the magnitude of such QG effects is sensitive to the change in average position of Bloch wavefunctions within the unit cell~\cite{ahn2021riemannian}. %\sout{as expressed in the change of electronic Bloch wavefunction layout within the unit cell upon excitation between bands \CL{Who do we cite here? SIPE?}.} \swch{I think we planned not to include any concrete statement about the connection between shift vector and displacement within the unit cell.} 
Previous works that studied shift current response in bilayer graphene and TMDs~\cite{cook2017design,xiong2021atomic,schankler2021large,ai20201t,xu2021colossal} predicted a strong effect due to their non-zero Berry curvature profile.

Quantum geometry-induced processes become more dominating in flat bands \cite{topp2021light,Xie2020}, where the large lattice constant sets the scale for the Berry connection in the flat bands. Recent pioneering studies considered twisted bilayer graphene at the magic angle (TBG)~\cite{kaplan2021,liu2020anomalous}, and confirmed the expectation of an unprecedented magnitude of the response.  

Our work expands these initial investigations, and provides a  systematic study of the shift current response on twist angle, filling factor, and encapsulation environment. We identify the role of band structure, relevant quantum geometry tensor elements, and the role of system's symmetries in determining the shift-current response. Particularly, we compute the shift-current response while including electron-electron interactions, and show that they significantly enhance the response as compared with a non-interacting model. Many recent works~\cite{guinea2018electrostatic,goodwinHartreeTheoryCalculations2020,ceaElectronicBandStructure2019,ceaBandStructureInsulating2020,rademakerChargeSmootheningBand2019,bultinckGroundStateHidden2020,AP19,SNP19,NadjPerge, choi2021correlation} have shown that interactions can also drastically alter the non-interacting bandstructure and associated wavefunctions profiles. These modifications, as we will show in this work, 
significantly affect the shift-current response studied in previous works~\cite{kaplan2021,liu2020anomalous} that considered a response of a non-interacting TBG only.  

 Inspired by recent experimental results~\cite{NadjPerge}, we consider specific types of electron-electron renormalizations of the electron bandstructure (see Fig.~\ref{Mainresultsfig}a-b,~\cite{guinea2018electrostatic,rademakerChargeSmootheningBand2019,goodwin2020hartree}) and demonstrate that these interactions can change both the magnitude and frequency response of the second-order conductivity. We argue that these changes arise from the interaction-induced band-flattening and modification of Bloch wavefunctions, specifically the quantum geometric connection, that are closely related to the shift-current photoresponse\cite{ PhysRevX.10.041041,ahn2021riemannian}.

  For simplicity, we carry our self-consistent calculations at temperature $T=0K$ but we expect the observed features to remain prominent up to liquid nitrogen temperatures, $T\approx77K$. This is because the characteristic energy scale for flat to dispersive band transitions that produces new resonances as well as the charge inhomogenity driven band-flattening are above that energy scale. Also, since  we are concerned with this high temperature regime, we do not consider the correlated behavior that typically emerges at temperatures $T\lesssim 15K$~\cite{cao1,cao2,Ilani19,Rozen2021}.

 In addition to the shift current, quantum geometry can also lead to other non-linear optical responses like injection current~\cite{Hosur2011,Chan2017,Juan2017} which arises from the change in group velocity of carriers upon excitation between two bands. However, for time-reversal symmetric systems such effects vanish for linearly polarized light~\cite{Ahn2020} and thus we ignore these effects in our present studies.

The paper is organized as follows: in section II we present a brief summary of our main results; in section III, we present the model used in our simulations, the mean-field treatment of Coulomb interactions and the methods employed to evaluate shift-current. We also compare different approaches used in literature and comment on their numerical amenability; in section IV, we proceed to study the shift-current response in a non-interacting twisted bilayer model and investigate the role of twist angle, sublattice offset, and symmetry properties. Additionally, we also analyse the contribution arising from different types of band transitions, e.g. flat to flat (FF) and flat to dispersive (FD) bands. We then try to understand the connection between observed shift-current and the real space profile of Bloch wavefunctions involved in transitions; in section V we discuss how these results are modified by interactions; finally we conclude by providing a summary of our analysis and specific experimental predictions.

\section{Summary of results}
\label{sec:summary}

We study the role of  twist angle,  doping, encapsulation environment, and electron-electron interactions on the shift-current response in twisted bilayer graphene. We find that in the absence of interactions, or equivalently at twist angles where non-interacting bandstructure accurately captures electronic properties, photoresponse has a universal form. This form is controlled by a moir\'e lengthscale and characteristic energies associated with flat-to-flat and flat-to-dispersive band transitions. The overall contribution of these two, flat-to-flat and flat-to-dispersive processes to the shift current also depends on the sublattice offset which can be tuned by varying the encapsulation environment. A finite sublattice offset leads to a gap opening between flat bands which can be controlled by the relative alignment between the graphe and hBN layer. Specifically, we find that sublattice offset does not drastically affect the gap between the flat and dispersive bands, unlike the gap between the flat bands. Therefore th sublattice offset allows
%, since the position of dispersive bands is not affected much by the substrate-induced sublattice offset, $\Delta$ and the flat bands are pushed away with increasing $\Delta$, 
 to control the relative importance of both types of transitions in shaping the photo-response.

%\textbf{[CAN WE EMPHASIZE THAT WE ARE THE FIRST TO CONSIDER THE EFFECTS OF INTERACTIONS? ALSO - say something like Perhaps most importantly] } \CL{How is this?}
Most importantly, electron-electron interactions significantly change the shift-current response as compared to a non-interacting syste [see Fig.~\ref{Mainresultsfig} (c-f)]. The role of interactions on the photoresponse becomes more pronounced as twist angle is brought closer to the magic angle, leading to a sharp increase in magnitude and narrowing of corresponding frequency window where resonances in shift-current were expected on the basis of non-interacting model. The key contribution of electron-electron interactions to the shift-current, however, is in altering photoresponse corresponding to transitions between flat and dispersive bands.  We attribute these features %on the basis of
to electron-electron interaction-driven changes to the band dispersion, the nature of Bloch wavefunctions and thus the resulting quantum geometry. 

Our results demonstrate that frequency range and magnitude can be tuned significantly by varying the twist angle and the substrate properties. Specifically, we observe a second-order conductivity of the order of $1000\mu A.nm/V^2$ in frequency range of 10-100 meV. This is in agreement with previous results of Ref.~\cite{kaplan2021} and \cite{xiong2021atomic} for TBG and gapped bilayer graphene respectively. We note however, that Ref.~\cite{kaplan2021} studies the frequency response in range 1-10meV and Ref.~\cite{xiong2021atomic} considers a frequency of 100meV. Finally, our work for the first time shows how photoresponse can serve as a probe of electron-electron interactions in TBG pointing towards a  novel experimental direction for the TBG field.

% \addCL{Both works focus on the photo-response of a non-interacting electron system.}
% \swch{not sure why do we need this line here? We mentioned about the previous works~\cite{kaplan2021} in the intro section. We can also mention it here if you like.}
%\swch{twist-angle made more prominent} \addswch{We find that the response of the system has a generic form near vicinity of the magic angle in the range $0.8^\circ<\theta<1.2^\circ$ and is dictated only by two energy scales: characteristic energy difference between same momentum states of the hole and electron flat bands; and the direct gap between flat and dispersive bands.} 

\section{Model and Methods}
\label{sec:model}
\subsection{TBG single-particle Hamiltonian}
\label{subsec:non_int_ham}

 \begin{figure*}[t] % If you use the onegrid enviroment then the sections start to get mixed up
 	\includegraphics[width=\linewidth]{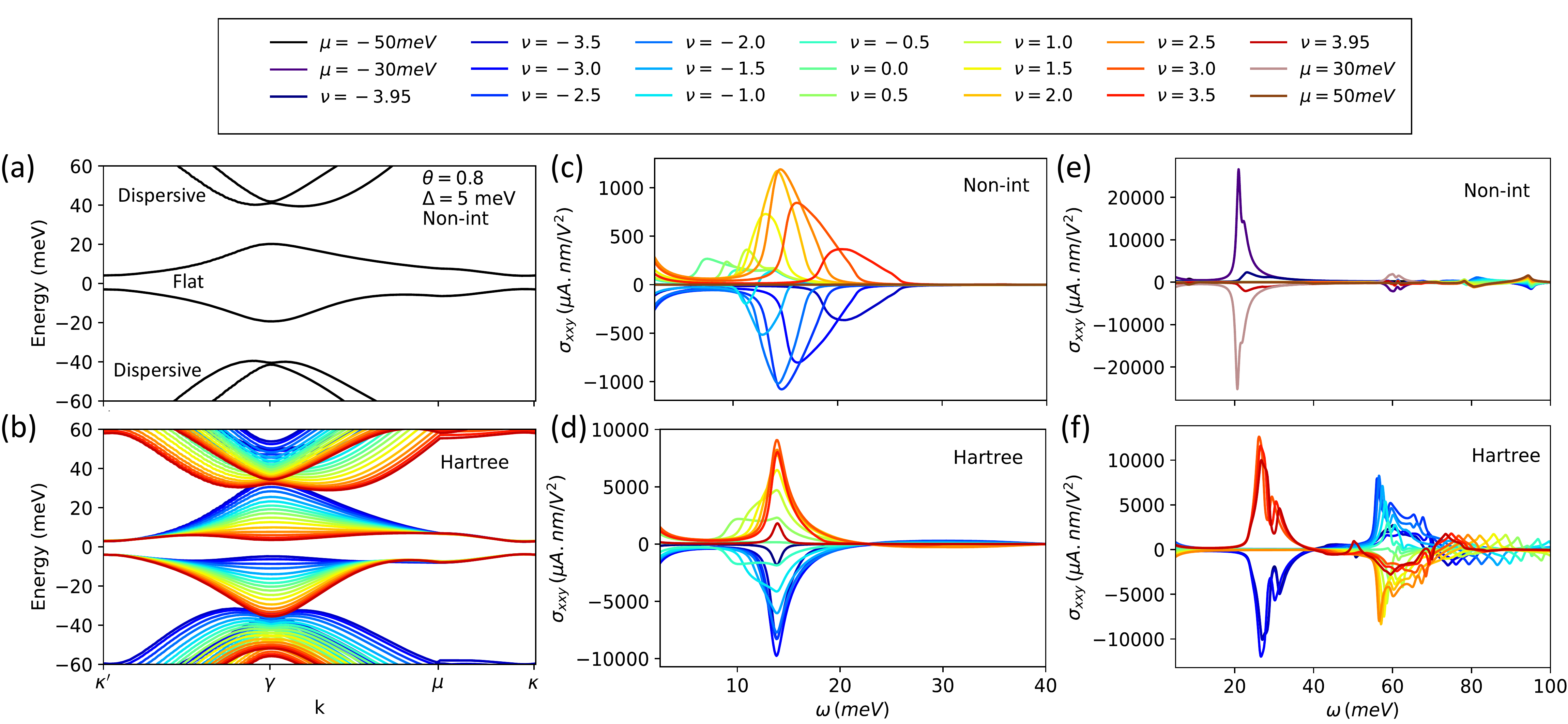}
 	
 	\caption{\textbf{Interaction-induced modifications in band structure and shift-current response of twisted bilayer graphene :} (a) Band structure for the non-interacting model presented in Eq.~\ref{eq:ham_cont}, (b) Band structure for the interacting case with Hartree corrections at different fillings, (c,d) contributions to second-order conductivity $\sigma_{xx}^{y}(0,\omega,-\omega)$ from flat-flat band transitions for non-interacting case and for the interacting case with Hartree corrections , and (e-f) contributions to second-order conductivity $\sigma_{xx}^{y}(0,\omega,-\omega)$ from flat-dispersive band transitions for the non-interacting case and for the interacting case with Hartree corrections. These Hartree corrections flatten both the flat and dispersive bands significantly as the filling is increased. This results in an enhanced second-order response and also gives rise to a second peak in flat-dispersive contribution.}
 	\label{Mainresultsfig}
 \end{figure*}

The single-particle energy spectrum of twisted bilayer graphene near the magic angle can be described with help of a continnuum model \cite{koshino2018,bistritzer2011moire,Santos2007}. Here, we follow the notation and model considered in Ref.~\cite{koshino2018} which gives a  Hamiltonian:
\begin{eqnarray}
\mathcal{H}_0 &=& \sum_{\gamma = \{\zeta,\sigma\}} \int_\Omega d^2\vec{r}~ \psi_\gamma^\dagger(\vec{r})  \hat{H}^{(\zeta,\sigma)} \psi_{\gamma}(\vec{r}),\\
\hat{H}^{(\zeta,\sigma)} &=& \begin{pmatrix} 
H_{\zeta 1}(\vec{r}) & U_{\zeta}^\dagger(\vec{r}) \\
U_{\zeta}(\vec{r}) & H_{\zeta 2}(\vec{r}) 
\end{pmatrix}\label{eq:ham_cont}
\end{eqnarray}
where $\Omega$ represents the moir\'e unit cell, $H_{\zeta, l}$ represents the intralayer Hamiltonian of layer $l=1,2$, and $U_{\zeta}(\vec{r})$ encodes the moir\'e interlayer hopping. The Hamiltonian is written in the basis of $(A_1, B_1, A_2, B_2)$ sites of the two layers and we use the shorthand notation, $\gamma\equiv\{\zeta(=\pm1),\sigma(=\pm1)\}$, for the valley and spin degrees of freedom, respectively. In rest of the paper, we refer to this Hamiltonian as the ``non-interacting model''.

The intralayer part of the Hamiltonian $H_{\zeta, l}$ is given by the two-dimensional Dirac equation expanded about the $\mathbf{K}^l_\zeta$ point of the original graphene layer,
\begin{equation}
H_l=-\hbar v\left[R(\pm\theta/2)(\mathbf{k}-\mathbf{K}_\zeta^l)\right]\cdot(\zeta \sigma_x,\sigma_y)+\Delta_l\sigma_z\,,
\end{equation}
where $\vec{k}$ is a momentum in the BZ of the original graphene layers,  $R\left(\pm\theta/2\right)$ is the $2\times 2$ two-dimensional matrix accounting for the rotation of layer $l=1(2)$ by an angle $\pm \theta/2$ about z-axis with respect to the initial AA stacked bilayer. We set $\hbar v/a = 2.1354$ eV as the kinetic energy scale for the Hamiltonians, $H_{\xi l}$ with $a=0.246$ nm being the original graphene's lattice constant. We also introduce a layer-dependent sublattice offset term, $\Delta_l\sigma_z$ that leads to a gap opening at the Dirac points.

The moir\'e interlayer potenial, $U_{\zeta}(\vec{r})$ in Eq.\eqref{eq:ham_cont}  can be approximated as:
\beq
\label{eq:app_moire_interlayer_coupling}
U_{\zeta}(\vec{r}) = \begin{pmatrix}
u & u'\\
u' & u\\\end{pmatrix}+\begin{pmatrix}
u & u'e^{-i2\pi \zeta/3}\\
u'e^{i2\pi \zeta/3} & u\\\end{pmatrix}e^{i\zeta \vec{G}_{1}^M\cdot\vec{r}} \nn\\+\begin{pmatrix}
u & u'e^{i2\pi \zeta/3}\\
u'e^{-i2\pi \zeta/3} & u\\\end{pmatrix}e^{i\zeta \left(\vec{G}_{1}^M +\vec{G}_{2}^M\right)\cdot\vec{r}}\,
\eeq
where we take  $u'=90meV$ and $u=0.4u'$ for twist angles near the magic angle. We justify our choice of parameters  in the next section. To diagonalize the Hamiltonian Eq.\eqref{eq:ham_cont} in $\vec{k}$-space, we can account for this interlayer potential by introducing a coupling between Bloch wave ansatzs at momentum $\mathbf{k}$ and $\mathbf{k}+\mathbf{G}$. Here, $\mathbf{G}=n_1\vec{G}_{1}^M+n_2 \vec{G}_2^M$ is a linear combination of moir\'e reciprocal vectors $\vec{G}_{1}^M$ and $\vec{G}_2^M$ where $n_1$ and $n_2$ are integers, and $G = |\vec{G}_{1}^M| = |\vec{G}_{2}^M|$ sets the characteristic momentum scale of the problem. These reciprocal lattice vectors are given by $\vec{G}_{i}^M=R(-\theta/2)\vec{G}_{i}-R(\theta/2)\vec{G}_{i}$ with $\vec{G}_{1}=(2\pi/a)\left(1,-1/\sqrt{3}\right)$ and $\vec{G}_{2}=(2\pi/a)\left(0,2/\sqrt{3}\right)$ being the reciprocal lattice vectors of a graphene monolayer.

\subsection{Mean-field interacting Hamiltonian}
\label{interactionsMF}
We consider electron-electron  interactions given by the Coulomb term:
\beq\label{eq:ham_int}
\mathcal{H}_c &=& \frac{1}{2} \int_\Omega d^2\vec{r}~d^2\vec{r}'~\delta\rho(\vec{r})~\V_{\mathrm{c}}(\vec{r}-\vec{r}')~\delta\rho(\vec{r}'),\\
\delta\rho(\vec{r}) &=& \sum_{\gamma = \{\zeta,\sigma\}} \psi_\gamma^\dagger(\vec{r})\psi_\gamma(\vec{r}) - \rho_{\mathrm{CN}}(\vec{r}),
\eeq
where $\delta \rho(\vec{r})$ is the density relative to that at charge neutrality, $\rho_{\mathrm{CN}}(\vec{r})$, and $\V_{\mathrm{c}}(\vec{r}-\vec{r}')$ is the Coulomb potential with a Fourier transform, $\V_{\mathrm{c}}(\vec{q})=2\pi e^2 / \epsilon q$. The dielectric constant $\epsilon$ depends on the substrate, and is treated as a free parameter (reasons to be made clear below). We approximate the above interaction term using a self-consistent Hartree approximation $\mathcal{H}_c \approx \mathcal{H}_H$ where
\begin{eqnarray}
    \mathcal{H}_H=\sum_{\gamma =\{\zeta,\sigma\}}\int_\Omega d^2\mathbf{r} V_H(\mathbf{r})\psi^\dagger_\gamma(\mathbf{r})\psi_\gamma(\mathbf{r})
\end{eqnarray}
with the Hartree potential
\begin{equation}
    V_H(\mathbf{r}) = \int_\Omega d^2\mathbf{r}'\V_{C}(\mathbf{r}-\mathbf{r}')\sum_\gamma \left\langle \psi_\gamma^\dagger(\mathbf{r}')\psi_\gamma(\mathbf{r}')\right\rangle_{H}
    \label{eq:Hartree}
\end{equation}
In the above expression $\la...\ra_H$ denotes a summation over occupied states measured from CNP ($\nu = 0$) \cite{guineaElectrostaticEffectsBand2018}.  When doping is increased with respect to the charge neutrality point, there is a preferential buildup 
 of charge at $AA$ sites in real space~\cite{guineaElectrostaticEffectsBand2018}, corresponding to electronic states near $\kappa$ points of the mini-Brillouin zone. The non-uniform spatial charge distribution generates an electrostatic potential that prefers an even redistribution of the electron density. In contrast, the real space charge distribution corresponding to electronic states near $\gamma$ point is more uniform in the unit cell. The effect of the electrostatic Hartree potential and the associated charge  redistribution thus leads to an increase in energy of the electronic states near the $\kappa$ and $\mu$ points compared to  the energy of states near the $\gamma$ point~\cite{guinea2018electrostatic,goodwinHartreeTheoryCalculations2020,rademakerChargeSmootheningBand2019}.

The effect of the Hartree potential becomes increasingly pronounced as a function of decreasing twist-angle, especially near the magic-angle where the non-interacting bandwidth is minimal. There is an increasing tendency towards band-inversion near the ${\gamma}$ point \cite{ceaElectronicBandStructure2019,goodwinHartreeTheoryCalculations2020}, a feature that has not been observed in experiments till date \cite{NadjPerge}.
However, it is important to note that other mechanisms, for example strain or a Fock term, can  act against this tendency towards band-inversion by increasing the overall bandwidth (both strain and Fock), or by contributing an opposing correction to the self-energy as compared to the Hartree term, Eq.~\eqref{eq:Hartree} (Fock only). In our analysis we focus only on the Hartree correction for a general $\theta$ and we omit results for $0.96^\circ < \theta  < 1.04^\circ$. In this range, we anticipate that the Hartree term would produce extreme band inversions not seen experimentally, which are most likely counteracted by another mechanism.

The bandstructure is obtained by employing the fitting procedure detailed in Ref.~\cite{NadjPerge}. The microscopic parameters of the Hamiltonian are determined by matching the theoretical energy spectrum of the system to the experimental STM results sufficiently far away from the magic-angle where no correlated effects are present. As explained in Ref.~\cite{NadjPerge}, for general agreement with the experimental results, it is necessary to use a dielectric constant $\epsilon$ larger than that set by the substrate. Similar procedures were employed in earlier studies \cite{xieWeakfieldHallResistivity2020,guineaElectrostaticEffectsBand2018,ceaElectronicBandStructure2019} and their origins theoretically can be justified by arguing that dispersive bands renormalize the dielectric constant for the Coulomb interaction projected to the flat-bands. The final renormalized bandstructures at fixed angle of $\theta=0.8^\circ$ is shown as a function of filling in Fig.~\ref{Mainresultsfig}(b). The most notable manifestation of the electron-electron interactions induced effects is the band-flattening around the  ${\gamma}$ and $\mu$ points beyond a certain filling.

We note that contribution of band-flattening effects on TBG properties were studied in recent works~\cite{klebl2020importance,lewandowski2021does,cea2021coulomb}. Qualitatively, the role of band-flattening was to either enhance the density of states at the Fermi level or to decrease overall bandwidth and as a result corresponding twist angle range over which correlated effects were expected, increased. We stress, however, that no other papers that studied NLOR in TBG\cite{kaplan2021,liu2020anomalous} have considered the role interactions can play in the photoresponse.
 
Before proceeding with the discussion of the shift-currents in TBG, we pause to clarify key assumptions of our modelling. Firstly we intentionally do not include the effects associated with the ``cascade transitions'' at integer fillings near magic-angle \cite{Ilani19,Yazdani19}, and the correlated effects such as superconductivity\cite{cao2} or insulating states\cite{cao1}. Physically this approximation is justified as optical NLOR experiments are typically performed at temperatures~\cite{wu2017giant} exceeding the characteristic temperatures ($T \lesssim 15 K$) associated with these phenomena~\cite{cao1,cao2,Ilani19,Rozen2021}. In principle however these effects, as well as more complex scenarios like the K-IVC state, could provide interesting constrains on and signatures in the photoresponse.  We expect Hartree corrections to persist to higher temperatures  as they are a reflection of charge inhomogenity of the system. Secondly we also neglect the possibility of varying interlayer hopping parameters $(u,u')$ in Eq.\eqref{eq:app_moire_interlayer_coupling}. We argue that this approximation is justified since our choice of $u\prime=90$ meV is comparable to typical literature values and the ratio of $\eta = u/u’ = 0.4$ is not too far from values quoted in literature that are typically in the range $\eta = 0.3$ to $0.7$.  Most crucially, however, even if $\eta$ were to be varied with the twist angle, the location of the van Hove singularity would remain fixed near filling of $\pm 1.9$ (or not drastically different energies) (see also Ref. \cite{qin2021critical}) until very high $\eta$ values of $0.8$. Such  values are typically not used in modelling. As such, we thus expect that although quantitative changes (such as precise frequency locations of peaks can vary) overall behavior of the system will remain qualitatively similar.

%second-order conductivity calculated in    \addCL{placing it in the experimentally interesting regime}\CL{Do you know of a reference for an experimentally interesting regime? If not we can cite theory shift-current papers that get comparable values.}.\CL{Include comment that frequencies are in tens of meV and numerical estimates to other works?}

\subsection{Shift current}
The shift current is a second-order DC response to an electromagnetic field arising from the interband optical excitations~\cite{Baltz1981}. In time-reversal symmetric systems, it depends on the linearly polarized component of light and its origin can be traced back to the real-space shift experienced by Bloch wavepacket upon excitation from the valence band to the conduction band. If the light is circularly polarized, then band transitions can also lead to an additional second-order DC response known as injection current which arises due to the change in carrier velocities upon excitation~\cite{sipe2000}. However, for a linearly polarized light, this kind of injection current response vanishes in a two-dimensional system if the time-reversal symmetry is preserved in the system. The shift current is sensitive to the  intraband and interband Berry connection of the bands involved in the transition process~\cite{sipe2000}, and hence offers a possibility to detect and harness the non-trivial band topology of Bloch bands in photovoltaic processes.

The shift-current response is determined by a rank three tensor, $\sigma^\mu_{\alpha\alpha}$ which satisfies
\begin{equation}
    \mathbf{J}^\mu=2\sigma^\mu_{\alpha\alpha}(0,\omega,-\omega)\mathcal{E}^\alpha(\omega)\mathcal{E}^\alpha(-\omega)
\end{equation}
where $\mathbf{J}^\mu$ is the $\mu^{\text{th}}$ component of current density, $\bm{\mathcal{E}}(t)=\bm{\mathcal{E}}(\omega)e^{i\omega t}+\bm{\mathcal{E}}(-\omega)e^{-i\omega t}$ is the electric field and greek indices denote spatial components, $\alpha = \{x,y\}$. The second-order conductivity tensor element, $\sigma^\mu_{\alpha\alpha}(0,\omega,-\omega)$  is given by (see Appendix~\ref{Appendix:shift_current} and Ref.~\cite{Fregoso2017}):
\begin{equation}
\sigma^{\mu}_{\alpha\alpha}(0,\omega,-\omega)=\frac{\pi e^3}{\hbar^2}\sum_{m,n}\int d^2\mathbf{k}f_{mn}|\mathbf{A}^\alpha_{mn}|^2\mathbf{S}_{mn}^{\mu\alpha}\delta(\omega-\epsilon_{mn})
\label{shiftvector}
\end{equation}
where  $\epsilon_{mn}=\epsilon_{m}-\epsilon_{n}$ is the energy difference between the two states that participate in the optical transition, and $f_{mn}=f_m-f_n$ is the difference in occupancy of their energy levels. The above expression features two geometric terms: a shift vector $\mathbf{S}_{mn}^{\mu\alpha}=\mathbf{A}^\mu_{mm}-\mathbf{A}^\mu_{nn}-\partial_\mu(\text{Arg}\mathbf{A}^\alpha_{mn})$ and the interband Berry connection $\mathbf{A}_{mn} = \frac{1}{i}\left<u_m|\nabla_\mathbf{k}|u_n\right>$ for  Bloch wavefunctions $\left|u_m\right>$ and $\left|u_m\right>$. This interband Berry connection enters into the shift vector expression as the EM field couples through dipole matrix and carries no other direct physical interpretation, whilst the shift-vector represents the shift experienced by the Bloch wavepacket upon excitation from  $m^\text{th}$ to $n^\text{th}$ band~\cite{belinicher1982kinetic,Fregoso2017,sturman2020ballistic}. We denote the integrand of the above expression as $R^{\alpha\alpha\mu}_{mn}=|\mathbf{A}^\alpha_{mn}|^2\mathbf{S}_{mn}^{\mu\alpha}$ and provided that the Hamiltonian has a linear dependence on momentum, it can also be expressed as
\begin{equation}
\begin{split}
R^{\alpha\alpha\mu}_{ab}=&\frac{1}{\epsilon_{ab}^2}\text{Im}\left[\frac{h_{ab}^\alpha h_{ba}^\mu \Delta_{ab}^\alpha}{\epsilon_{ab}}\right]+\\& \frac{1}{\epsilon_{ab}^2}\text{Im}\left[\sum_{d\ne a,b}\left(\frac{h_{ba}^\alpha h_{ad}^\mu h_{db}^\alpha}{\epsilon_{ad}}-\frac{h_{ba}^\alpha h_{db}^\mu h_{ad}^\alpha}{\epsilon_{db}}\right)\right].
\end{split}\
\label{sumrule1}
\end{equation}
In the above expression we introduced a shorthand notation $h^{\alpha}_{ab}=\left<a|\nabla_{k_\alpha}H|b\right>$, $h^{\alpha\beta}_{ab}=\left<a|\nabla_{k_\alpha}\nabla_{k_\beta}H|b\right>$ for the momentum derivatives of the Hamiltonian. The above expression for shift-current is equivalent to the sum rule commonly used to calculate the shift vector~\cite{Fregoso2017}. We stress that if the time-reversal symmetry is broken intrinsically or by application of circularly polarized light, there can be an additional contribution to the current density which is linear in scattering time and is  known as the injection current~\cite{sipe2000,Hosur2011,Chan2017,Juan2017}.

In literature there are several methods to calculate second-order NLOR conductivity \cite{sipe2000,Parker19,cook2017design,zhang2018}. In fact, in some previous works Eq.\eqref{shiftvector} is often presented in a slightly different form without any explicit reference to the shift-vector. For example, one of the most common expressions\cite{zhang2018, liu2020anomalous} is of the form
    \begin{align}
        \sigma^{\mu}_{\alpha\beta}&=-\frac{e^3}{\hbar^2\omega^2}\times\nonumber\\
       &\text{Re}\left(\sum_{\Omega=\pm\omega,{m,n,l}}\int d^2\mathbf{k} \frac{h_{nl}^\alpha h_{lm}^\beta h_{mn}^\mu}{(\varepsilon_{mn}-i\eta)(\varepsilon_{nl}+\Omega-i\eta)}\right)
    \end{align}
which we show in the Appendix~\ref{Appendix:shift_current} is equivalent to Eq.~\ref{shiftvector} except for the injection current term which arises for $m=n$ in the above summation. This injection current term vanishes if $\alpha=\beta$ or if the TRS is preserved. We also note that although at first glance the above expression incorporates three states involved in the transition process, whilst the expression of Eq.\eqref{shiftvector} features only two. This is resolved by realising that one of the states in the above expression comes from a virtual transition and is accounted for explicitly in the summation as shown in Eq.\eqref{sumrule1}.
    
The above shift vector expression of Eq.\eqref{shiftvector} can be calculated directly from the Berry connection matrix. In practice, however, the direct numerical approach is plagued by gauge fixing issues and hence is not reliable. We instead consider the approach used by Ref.~\cite{Parker19,cook2017design} to calculate the shift vector using the sum rule described in Eq.~\ref{sumrule1} which, as explained above, is equivalent to the three velocity expression in Ref.~\cite{Parker19}. This approach is more amenable for numerical simulations and also puts different expressions considered above on an equal footing. We provide a detailed derivation of the shift-current conductivity in Appendix~\ref{Appendix:shift_current}, and elucidate the connection between different expressions encountered in the literature.

\subsection{Symmetry constraints on second-order conductivity}
\label{subsec:symm}
It is  well-known that  second-order optical processes are observed only in non-centrosymmetric materials~\cite{belinicher1980photogalvanic,sturman2020ballistic}. The number of non-vanishing and independent elements of the second-order conductivity tensor can be deduced directly from the symmetry groups of the crystal via a simple application of group theory.  The continuum TBG model considered here has $D_{3}$ symmetry generated by a $C_{3z}$ and $C_{2y}$ when the sublattice offset term is the same on both layers. However, when $\Delta_1\ne\Delta_2$, the symmetry group reduces to $C_{3z}$. As a result of these symmetry properties, as derived in Appendix~\ref{Appendix:Symm}, we expect the conductivity tensor to satisfy
\begin{equation}
\begin{split}
\sigma_{xx}^y=-\sigma_{yy}^y=\sigma_{xy}^y=\sigma_{yx}^x\ne 0\\
\sigma_{yy}^x=-\sigma_{xx}^x=\sigma_{yx}^y=\sigma_{xy}^y=0.
\end{split}
\end{equation} 
when $\Delta_1=\Delta_2$. On the other hand, for $\Delta_1\ne\Delta_2$, we have
\begin{equation}
\begin{split}
\sigma_{xx}^y=-\sigma_{yy}^y=\sigma_{xy}^y=\sigma_{yx}^x\ne 0\\
\sigma_{yy}^x=-\sigma_{xx}^x=\sigma_{yx}^y=\sigma_{xy}^y\ne 0.
\end{split}
\end{equation}
Indeed these group theory based conclusion can be explicitly checked through evaluation of Eq.\eqref{shiftvector}.

 \begin{figure*}[t]
 	\includegraphics[width=\linewidth]{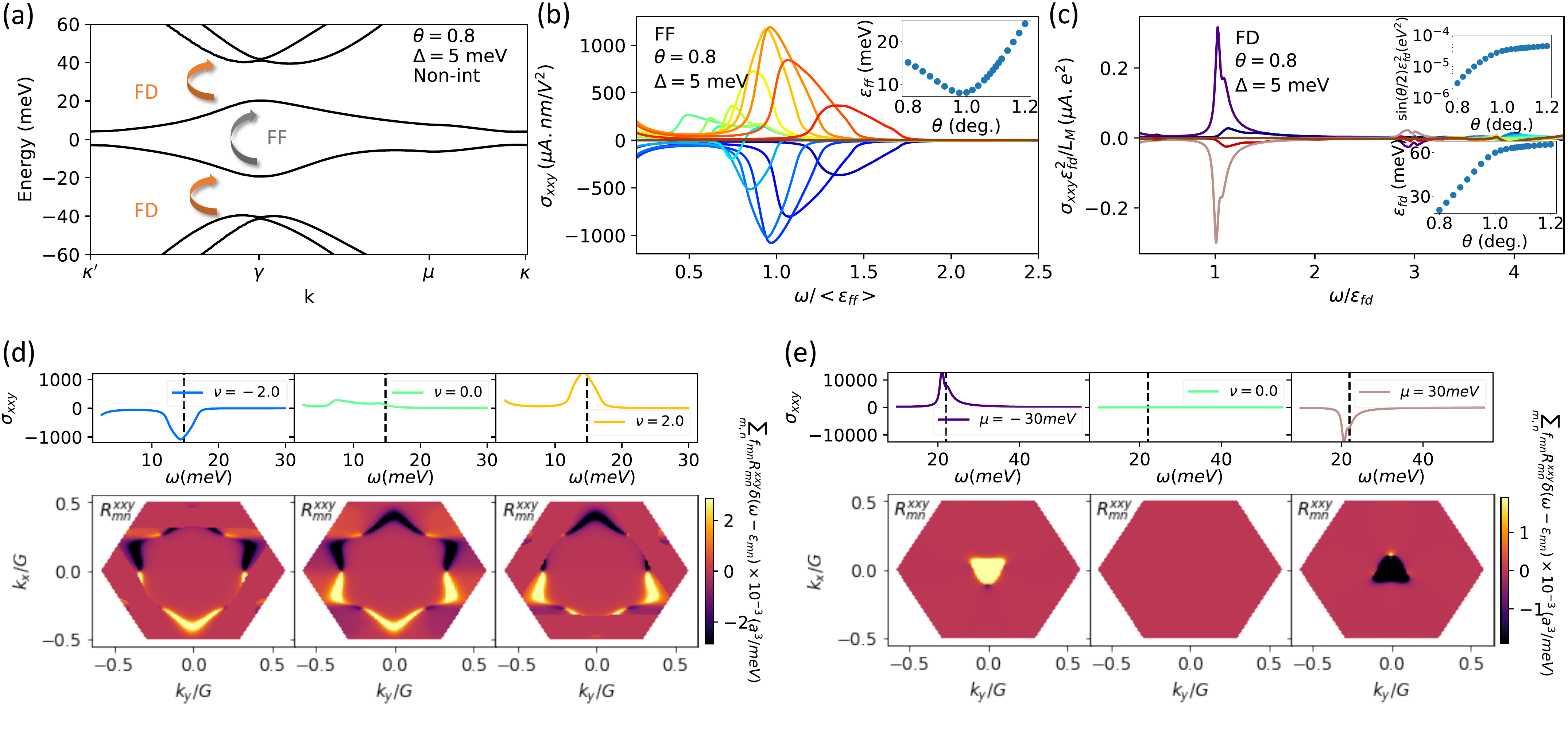}
 	\caption{\textbf{Band structure and shift-current response for twisted bilayer graphene near magic angle :} (a) Bandstructure for non-interacting twisted bilayer graphene for twist angle $\theta=0.8^{\circ}$ and sublattice offset $\Delta=5meV$ on both layers. Here, FF and FD denote flat-to-flat and flat-to-dispersive band transitions, respectively. (b) FF contribution to second-order conductivity as a function of frequency shown in units of average gap between two flat bands. The behavior of this average gap with twist angle $\theta$ is shown in the inset. Note that the van Hove singularity in our model occurs near $\nu\approx \pm 1.9$, and thus the peak value of $\sigma_{xxy}$ is significantly large for fillings close to this value. (c) FD contribution to shift current conductivity scaled by $\varepsilon_{fd}^2/L_M$ (see main text for justification) as a function of frequency in units of the band gap between flat and dispersive bands denoted by $\varepsilon_{fd}$. Dependence of this energy gap $\varepsilon_{fd}$ and the moir\'e length dependent parameter $\varepsilon_{fd}^2/L_M$ is shown in insets. (d,e) The top row shows the conductivity at different fillings and the bottom row shows $k$ space profile of the integrand $\sum_{m,n}f_{mn}R^{xxy}_{mn}\delta(\omega-\varepsilon_{mn})$ contributing to the second order conductivity at the frequency corresponding to the dashed line for flat-flat and flat-dispersive transitions shown in the upper panel, respectively.}
 	\label{nonintresults}
 \end{figure*}  

\section{Shift current in the non-interacting case}
\label{sec:NIshiftcurrent}
In this section, we investigate how this shift current changes with different parameters of the system, for now, in the absence of electron-electron interaction. We identify two different contributions to the photocurrent in presence of linearly polarized light - (1) those originating from transitions from a flat band to another flat band which is referred as FF, and (2) arising due to transitions between a flat band and a dispersive band which is referred as FD contribution. Both are schematically depicted in Fig.~\ref{nonintresults} (a).

The frequency dependence of the second-order conductivity is dictated by the integrand, %$\sum_{m,n}f_{mn}R^{\alpha\alpha\mu}_{mn}\delta(\omega-\varepsilon_{mn})$ 
~$\sum_{m,n}f_{mn}R^{\alpha\alpha\mu}_{mn}\delta(\omega-\varepsilon_{mn})$. For the cases of flat-flat transitions, as expected, the peak frequency is close to the average gap between two flat bands as shown in Fig.~\ref{nonintresults}~(c) and the obtained second-order conductivity looks almost identical for all twist angles away from the magic angle. This result is to be expected as the TBG continuum model (near the magic-angle) for various $\theta$ produces qualitatively similar bandstructures (and wavefunctions) up to an overall scale factor. Moreover, the peak of the response in Fig.~\ref{nonintresults}~(c) occurs near $\nu\approx 2$, which corresponds to the filling associated with the van Hove singularity location $\nu\approx 1.9$ for our continuum model parameters. As $(u,u')$ parameters of Eq.\eqref{eq:ham_cont} are kept constant for all $\theta$ this location of the van Hove singularity will remain at the same filling further demonstrating the similarity of the response for a wide range of angles. 

In addition to being sensitive to the energy gap between flat bands, the overall behavior of second-order conductivity also depends on the $k$ space profile of the quantity $\sum_{m,n}f_{mn}R^{\alpha\alpha\mu}_{mn}\delta(\omega-\varepsilon_{mn})$ which we refer to as shift-vector integrand.  An important point to notice is the profile of shift vector integrand, $R^{\alpha\alpha\mu}_{mn}$ in momentum space peaks around Dirac points and has equal regions of positive and negative values as shown in the middle panel of Fig.~\ref{FFHartree}(d). However, the integral in Eq.~\ref{shiftvector} also has a $\delta(\omega-\varepsilon_{mn})$ factor and the energy contours for a given valley are not symmetric about $k_x=0$

which results in a large net contribution whenever the filling is non-zero as shown in Fig.~\ref{nonintresults}~(d) (left and right panels). This imbalance between positive and negative regions is more prominent for the fillings, $\nu\approx2$ which results in a significant contribution from the regions near $\mu$ point (Fig.~\ref{fig:compareFFkspace}). It is worth noticing that these regions are extremely flat (as they lie in vicinity of the van Hove singularity) and thus contribute heavily due to large density of states. At the same time, the shift vector from $\zeta = -1$ valley is opposite of the $\zeta = +1$ valley but at the same time energy contours are  time-reversal partners of each other which results in the same contribution to second-order conductivity. We can apply similar arguments to conclude that the contribution from shift vector $R^{yyx}$ would vanish as the energy contours are symmetric about $k_y=0$ but $R^{yyx}(k_x,k_y)=-R^{yyx}(k_x,-k_y)$. This is of course to be expected from the symmetry analysis of the Sec. \ref{subsec:symm}, but here is demonstrated as an explicit consequence of the integrand $R^{\alpha\alpha\mu}$.

In fact, the frequency dependence of the photoresponse for the non-interacting model is largely decided by the gap, which can be tuned by improving the lattice alignment between between hBN layers and tBLG sample. In our plots we considered a sublattice offset $\Delta=5meV$ for both layers which results in a gap of about $10 meV$ and thus the contribution from flat-flat bands peak around $10meV$. We present our results for other sublattice offset in Fig.~\ref{Fig_Deltachange} of the Supplemental Materials. As expected, we notice that the frequency response can be tuned by varying the sublattice offset. However, as we keep on increasing $\Delta$, we notice that the second-order response starts to diminish. This is to be expected as in addition to a suppression coming from the energy denominators exemplified in Eq.\eqref{sumrule1}, the wave-function overlap between the bands decreases as well as bands become more decoupled with increasing $\Delta$. This also suppresses the band topology.

An important point to notice about the flat-flat contribution is that it indirectly depends on the presence of dispersive bands. The shift vector between two-flat bands has contribution from virtual transitions to dispersive bands as evident from  the second term in Eq.~\ref{sumrule1} even though we are focusing here on direct flat to flat transitions. As a result, the number of dispersive bands also play an important role in deciding the behavior of shift-vector between flat-flat bands.  In our simulation we found that it was necessary to include  ten dispersive bands while evaluating this shift vector using the expression Eq.~\ref{sumrule1} (where virtual transitions are captured by the second term) to achieve convergence of the second-order conductivity.

We now focus on  another contribution to second-order conductivity that comes from real transitions between a flat band and a dispersive band depicted by orange arrows in Fig~\ref{nonintresults} (a) (which we refer to as FD). In this case, the integrand $
\sum_{mn}f_{mn}R^{xxy}_{mn}\delta(\omega-\varepsilon_{mn})$ (we include those indices which account for transitions between flat and dispersive bands) is concentrated around the $\gamma$ point in $k$ space (Fig.~\ref{nonintresults}~(e)), and thus we observe a significant non-zero contribution only when the Fermi level lies between a flat band and a dispersive band.  

Just as in the case of a flat-to-flat response of Fig.\ref{nonintresults}(b), we can similarly extract $\theta$ independent form of the photoresponse corresponding to flat-to-dispersive transitions. As expected from Eq~\ref{sumrule1}, the integrand $R^{xxy}_{mn}$ decreases as $1/\varepsilon_{fd}^2$ where $\varepsilon_{fd}$ is the gap between the flat band and the dispersive band. This gap shows a very strong dependence on twist angle $\theta$ as it increases sharply with the increase in mini-BZ size. The integral also carries an additional length-scale dependence. Hence, to present results  in a $\theta$ independent manner,  we rescale the response by a prefactor $\epsilon_{fd}^2/L_M$, where $L_M$ is the moir\'e length as shown in Fig.~\ref{nonintresults} (c). Although the main plot shown in this figure was obtained for twist angle $\theta=0.8^\circ$, it looks quantitatively identical for all other twist angles near the magic angle. As evident from the behavior of the scaling factor $\epsilon_{fd}^2/L_M$, the second-order conductivity is orders of magnitude larger for $\theta=0.8^\circ$ in comparison to $\theta>1^\circ$, and the peak value is roughly equal to $20000\mu A.nm/V^2$ for this twist angle which is an order of magnitude higher than that corresponding to flat-to-flat transitions of Fig.\ref{nonintresults}(b). We also highlight that the largest response is seen at frequency corresponding to that of a flat to dispersive band gap, $\omega = \epsilon_{fd}$, but additional resonances occur at higher frequencies. We will explore these features in the following section.

Similar to the first contribution to shift-current response, the second contribution arising from the flat-to-dispersive band transitions is also influenced by the substrate properties.
When the sublattice offset, $\Delta$ is increased from $5$ mev to $10$ meV, we notice that the FD signal is shifted to a lower frequency and the peak becomes more pronounced as shown in the Supplemental Fig.~\ref{Fig_Deltachange}(c). This can be explained on the basis of the shift in band energies (Fig.~\ref{Fig_Deltachange}(a)). An increased $\Delta$ increases the gap between flat bands but does not affect the dispersive bands much. As a result, the gap between flat and dispersive bands starts to decrease. A smaller value of the gap, $\varepsilon_{fd}$, shifts the peak to lower frequency and also increases the value of integrand which scales as $1/\varepsilon_{fd}$ as mentioned earlier. However, if we increase the sublattice offset further, it suppresses the overlap between Bloch wavefunctions as discussed previously in the context of flat-to-flat transitions  and the shift current signal is diminished as shown in Fig.~\ref{Fig_Deltachange}. This shows that the sublattice offset can serve as an important knob to tune the optical response. Additionally, the direction of current density and its relation to the polarization of EM field can also be modified by changing the sublattice offset independently in two-layers. As discussed in Sec.~\ref{subsec:symm}, the constraints on the second-order conductivity tensor are different for $\Delta_1=\Delta_2$ case and $\Delta_1\ne\Delta_2$ case. Here, in Figs.~\ref{nonintresults}-\ref{FDHartree}, we have considered $\Delta_1=\Delta_2$, and thus the only non-zero components are $\sigma_{xx}^y,\sigma_{yy}^y,\sigma_{xy}^y,\sigma_{yx}^x\ne 0$ which can all be expressed in terms of $\sigma_{xx}^y$ plotted in these figures. We also verified the relation between different elements as shown in Fig.~\ref{fig:differentdelta}.  However, for $\Delta_1\ne\Delta_2$, there are two independent non-zero elements which are shown in the lower panel of the same figure.

\section{Effects of Interactions on shift-current response}

Next, we  discuss how the shift current response is modified by electron-electron interactions which we incorporate by using mean-field methods described in section~\ref{interactionsMF}. As shown in Fig.~\ref{Mainresultsfig} (b), one of the most prominent effect of interactions is the band-flattening of the flat bands near the $\gamma$ point causing a large enhancement of density of states. Additionally, these interactions also affect the structure of the Bloch wavefunction in real-space which modified the shift-vector.

\begin{figure}[h]
 	\includegraphics[width=\linewidth]{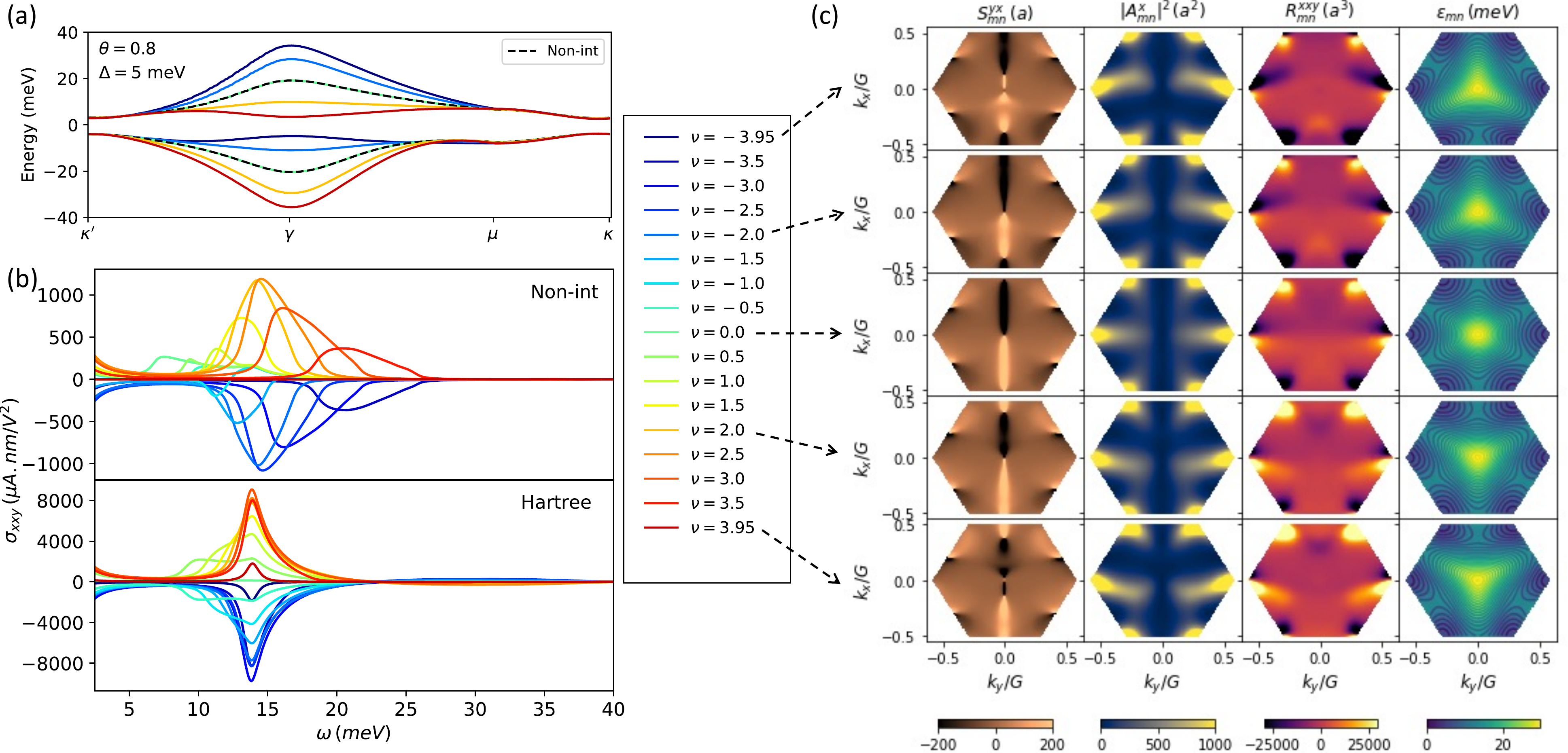}
 		\caption{\textbf{Interaction induced modification of band structure, second-order conducitivity, shift-vector and interband Berry connection for flat-to-flat band transitions.} (a) Band structure with Hartree corrections. All curves are shifted to the same energy at $\kappa'$. Note how as filling increases (decreases), electron (hole) flat band flattens and hole (electron) band broadens. (b) FF contribution to shift-current conductivity for the non-interacting case (upper panel) and the interacting case (bottom panel). Electron-electron band flattening increases the overall magnitude of the response and narrows the resonance in frequency. (c) $k$ space profiles of the shiftvector ($S^{yx}_{mn}$) in units of the lattice constant of the monolayer graphene lattice, interband Berry connection magnitude square, $|A^x_{mn}|^2$, the integrand $R^{xxy}_{mn}$ and energy contours for $\varepsilon_{fd}$ in $k$ space for the transition between two flat-flat bands at four different fillings used to calculate shift current response in Eq.~\ref{shiftvector}. It is worth noticing that the shift vector can be orders of magnitude larger than the lattice constant, $a$. This is expected as the Berry connection is roughly of the order of the lattice constant of moir\'e lattice.}
 	\label{FFHartree}
 \end{figure}

\begin{figure}
    \centering
    \includegraphics[width=\linewidth]{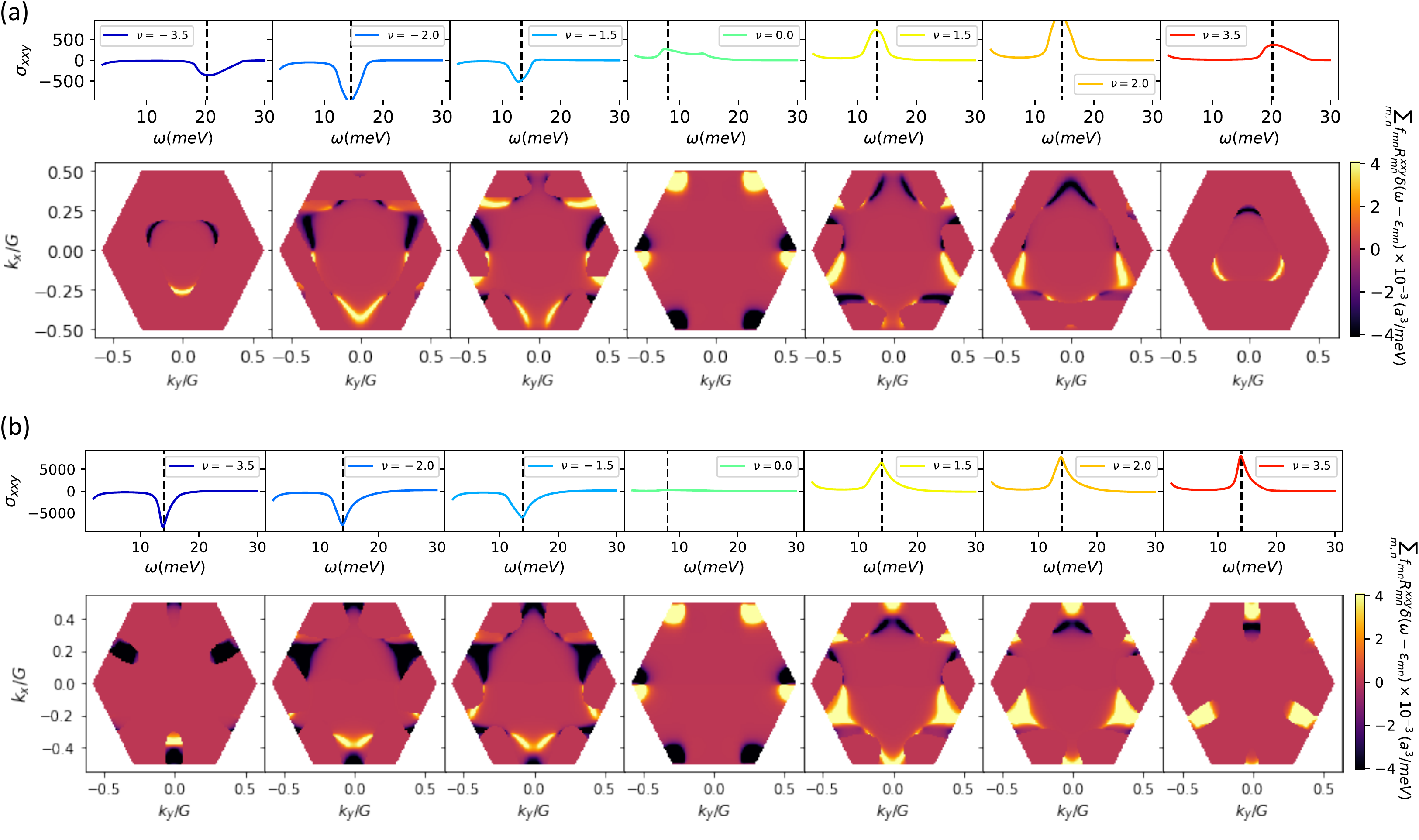}
    \caption{\textbf{Comparison between the momentum space profile of FF contribution of shift-current conductivity for the non-interacting and interacting case at peak frequencies for different filling factors.} The flat-to-flat band transition contribution to the peak of second-order conductivity from different $k$ points within the mini BZ at different filling factors for (a) non-interacting model and (b) interacting model with Hartree corrections. In each subfigure, the upper panel shows the variation of shift-current conductivity with frequency at a given filling and the lower panel shows the $k$ space profile of the shift-current integrand from Eq.~\ref{shiftvector} for the flat-to-flat band transitions at frequencies corresponding to the dashed line in the upper panel. We notice a significant increase in the contribution from the regions near the $\mu$ point which mainly arises from the  band-flattening effect of interactions.}
    \label{fig:compareFFkspace}
\end{figure}

For the flat-flat contribution shown in non-interacting case, we noticed that the $\sigma_{xx}^{y}$ peak was significantly larger for fillings, $\nu\approx2$ Fig.~\ref{nonintresults}(b). We explained this behavior on the basis of a significant contribution from the extreme flat regions around $\mu$ points, corresponding to the van Hove singularities, as depicted in Fig.~\ref{nonintresults}(b) and Fig.~\ref{fig:compareFFkspace}(a).  Upon increasing the filling further beyond these flat-regions, the transitions to these states was Pauli blocked and they no longer contributed to the optical response in non-interacting case. However, when electron-electron interactions are included in the analysis, we notice that these flat-regions around $\mu$ point expand further in $k$ space as shown in Fig.~\ref{FFHartree}(a) until they span the whole mini-BZ (when the $\gamma$ point is locally flat). Now, these extremely flat regions can participate in band transitions even at much larger fillings. It consequently affects the peaks at larger fillings, e.g $|\nu|>3$, which not only increase in strength but also shift in frequency and coincide with the peaks at fillings $|\nu|=1.5,2$. This behavior clearly arises due to the increased density of states coming from Hartree band-flattening that shifts van Hove singularity to higher fillings. Additionally, we also notice a change in the profile of flat-dispersive contribution of the integrand $\sum_{mn}R_{mn}^{xxy}\delta(\omega-\varepsilon_{mn})$ along $\gamma-\mu$ line which leads to an increased asymmetry in positive and negative regions of the mini-BZ with increasing $|\nu|$ as depicted in the third column of Fig.~\ref{FFHartree}(c), and hence an enhanced response.

\begin{figure}[h]
 	\includegraphics[width=\linewidth]{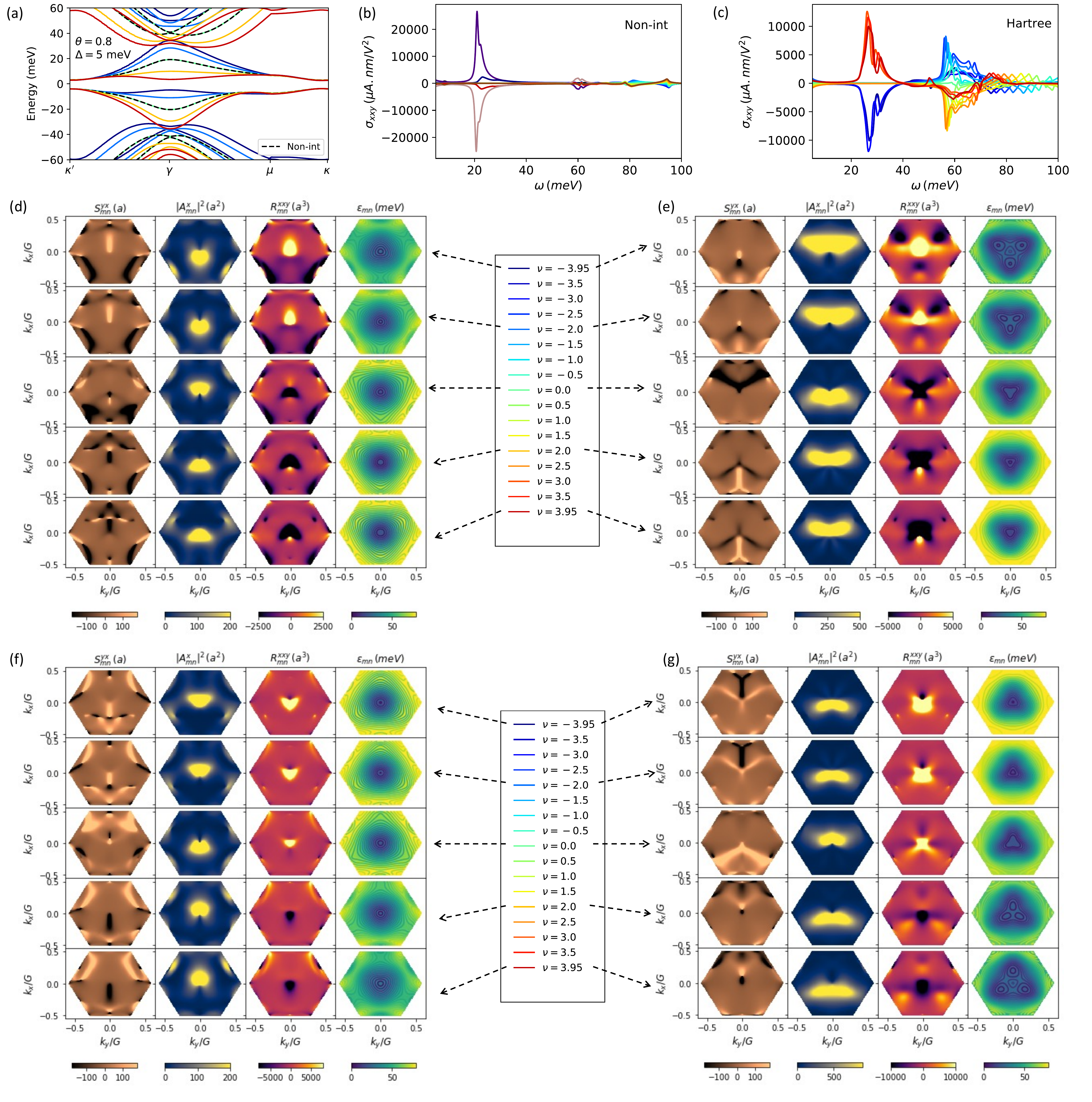}
 	\caption{\textbf{Interaction induced modification of band structure, second-order conducitivity, shift-vector and interband Berry connection for flat-to-dispersive band transitions.} (a) Band structure with Hartree corrections showing flat and dispersive bands, (b) FD contribution to shift-current conductivity for the non-interacting case, (c) FD contribution when Hartree corrections are included. Note the appearance of additional peaks in (c) as compared to (b). (d-f) Shiftvector ($S^{yx}_{mn}$), Interband Berry connection magnitude square, $|A^x_{mn}|^2$, the integrand $R^{xxy}_{mn}$ and energy contours for $\varepsilon_{fd}$ in $k$ space for the four FD transitions where (d-e) represents transitions between the hole flat band and hole dispersive bands, and (e-f) describe transitions between the electron flat band and the electron dispersive bands.}
 	\label{FDHartree}
 \end{figure}

These interaction-induced changes in band structure also affect the contribution coming from transitions between flat and dispersive bands. One obvious modification arises from the changes in band structure which are quite prominent around $\gamma$ point. This region was the hotspot for FD contribution in non-interacting case as discussed in Sec.~\ref{sec:NIshiftcurrent} and shown in Fig.~\ref{nonintresults}(e). The Hartree corrections to the non-interacting Hamiltonian increases the gap at $\gamma$ points. It also results in an increased band flattening of dispersive bands, which in turn decreases the gap significantly in a large region of mini BZ around $\mu$ points as shown in Fig~\ref{FDHartree} (a). 

\begin{figure}
    \centering
    \includegraphics[width=\linewidth]{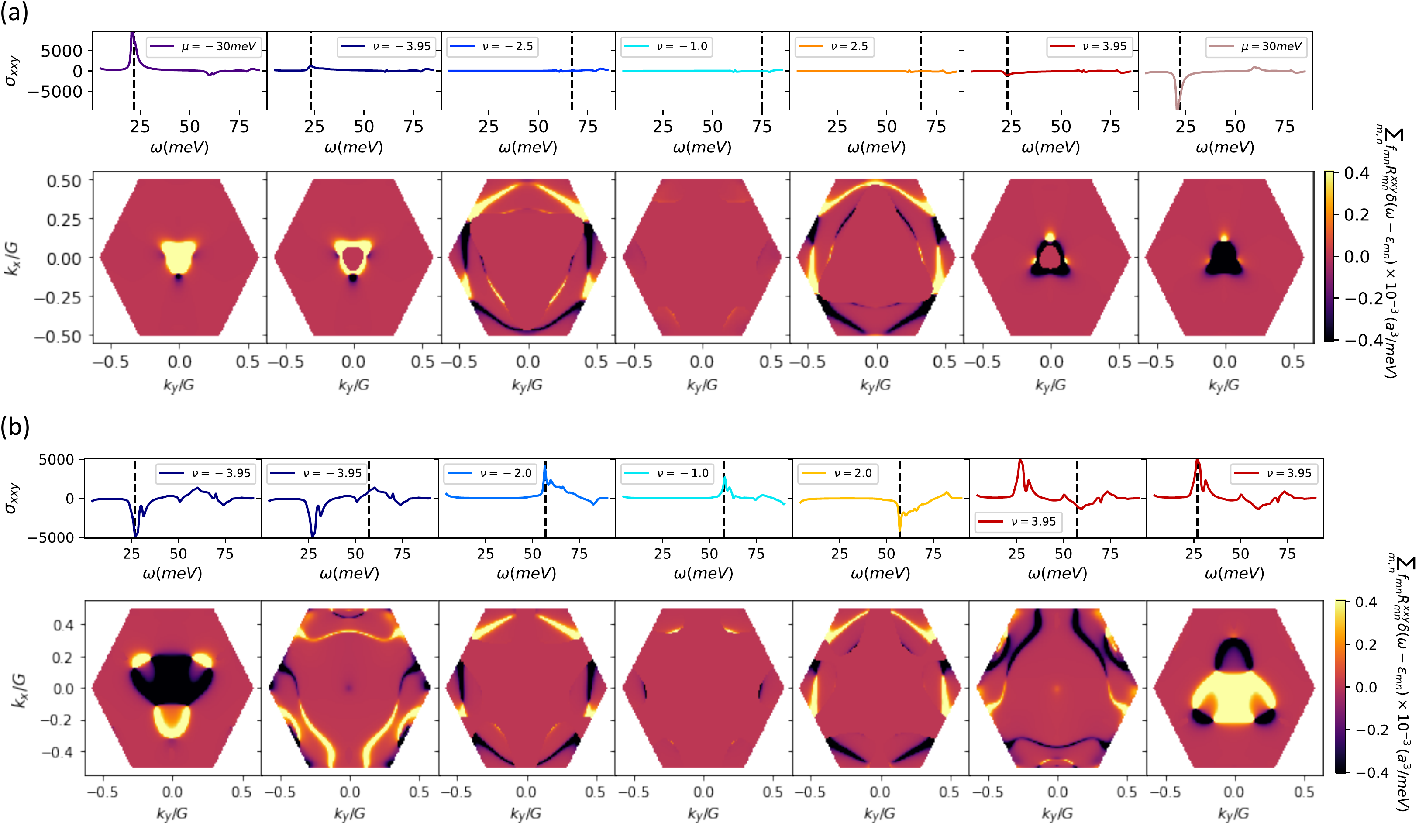}
    \caption{\textbf{Comparison between the momentum space profile of FD contribution of shift-current conductivity for the non-interacting and interacting case at peak frequencies for different filling factors.} The flat-to-dispersive band transition contribution to the peak of second-order conductivity from different $k$ points within the mini BZ at different filling factors for (a) non-interacting model and (b) interacting model with Hartree corrections. In each subfigure, the upper panel shows the variation of shift-current conductivity with frequency at a given filling and the lower panel shows the $k$ space profile of the shift-current integrand from Eq.~\ref{shiftvector} for the flat-to-dispersive bands transitions at frequencies corresponding to the dashed line in the upper panel. We notice a significant increase in the contribution from the regions near the $\mu$ point which mainly arises from the  band-flattening effect of interactions.}
    \label{fig:compareFDkspace}
\end{figure}

These Hartree corrections to the band structure and Bloch wavefunctions also modify the shift current integrand, $R^{xxy}$. Its momentum profile exhibits a significant increase in regions away from $\gamma$ point as shown in Fig.~\ref{FDHartree} (d-g) and Fig.~\ref{fig:compareFDkspace} (b). These factors give rise to some unexpected features in the second-order conductivity response. We can now observe a reasonably large contribution at fillings less than $|\nu|=4$, which arises due to the spreading of $R^{xxy}_{mn}$ in mini BZ as shown in the bottom panel of Fig.\ref{fig:compareFDkspace}(b).

Arguably, however, the most important role (at least experimentally) of these filling-dependent corrections is the appearance of new features in the second-order shift current conductivity. Specifically, there is also a second peak, in Fig.\ref{FDHartree}(c) at $\omega \approx 60$ meV, which has the opposite sign, to the peak at $\omega \approx 25$ meV. We attribute this second-peak to  transitions that involve van-Hove singularity points of the flat-band as their frequency is quite close to the energy gap around those $k$ points. This is further substantiated by the fact that the integrand in these regions is opposite to the that of the contribution from the $\gamma$ point as shown in Fig.~\ref{fig:compareFDkspace}(b) and the third column of Fig~\ref{FDHartree} (e,f). We argue that this enhanced response and the appearance of the second peak can act as a probe of interaction-induced changes to both the band structure and the quantum geometry. Most crucially, however, this additional peak occurs at frequencies that far exceed those characteristic frequencies of flat-to-flat band transitions (few meV's), placing it more firmly in the characteristic range of optical experiments (tens of meV's).

\section{Discussion}

In this work, we  presented a detailed analysis of the shift-current response in TBG, and investigated  the role of twist angle, doping, encapsulation environment, and interactions on the shift-current response of twisted bilayer graphene. We identified two different contributions: one arising from the transitions between two flat bands and another from the transitions between a flat band and a dispersive band as shown in Fig.\ref{Mainresultsfig}.  In the absence of interactions, the first and second contributions resulted in a second-order conductivity with peak values of ${\sim } 1000\,\mu A.nm/V^2$ and ${\sim } 10000\,\mu A.nm/V^2$, respectively with the typical frequency dependence tunable by changing the twist angle and the sublattice offset. This giant photoresponse arising from the non-trivial band topology of flat-bands in TBG renders it an exceptional material for photovoltaic applications in THz range. Additionally, we showed that interactions can significantly alter the photoresponse of TBG. This opens up a novel route to probe interaction-induced changes to band structure and quantum geometry with the help of optical probes.

Alongside the shift current response, CPGE (Circular photogalvanic effect) and, generally, injection photocurrents, also occur in materials with Dirac cone dispersions. The injection current, emerges from the difference in group velocities between the original and excited bands, and is proportional to the electronic relaxation time. It is usually the dominant second-order photocurrent response in Weyl semi-metals ~\cite{Juan2017,Chan2017}. It requires, however, circularly polarized light or tilted Dirac cones illuminated by linearly polarized light to be non-zero. Our work considered only the linear polarization response, and ignored these additional terms which we expect are either subleading in TBG or vanish by symmetry considerations. Specifically, for a two-dimensional system with $C_{3z}$ symmetry, even the circular polarization cannot generate an in-plane injection current at normal incidence~ (see \cite{dresselhaus2007group} and discussion in appendix \ref{note2}). As such, unless $C_{3z}$ symmetry is lifted, for example by applying a strain (as recently shown in Ref. \cite{2021arXiv210707526A}), we expect injection current to vanish under these conditions in TBG.

%Shift-vector part e
Another interesting aspect is the dependence of the shift-vector on the nature of the perturbation, i.e., the momentum-derivative of the excitation matrix phase~\cite{Chaudhary2018}. It could be interesting to contrast this contribution to the shift current with currents induced by other non-equilibrium perturbations arising from coupling between EM fields and other degrees of freedom such as orbital or phononic degrees of freedom.

Furthermore, in this manuscript we mainly focused on the photoresponse originating from interband processes. However, if the spatial symmetry of the system is lowered further by breaking some mirror symmetries, we could also get a second-order contribution from intraband processes which are captured by Berry curvature dipole~\cite{sodemann2015quantum}. Such processes can be made to contribute to the non-linear optical response by applying a strain as discussed in Ref.~\cite{Raffaele2019,Pantalean2021}. In TBG, we expect the strain-induced contribution to be of the same order of magnitude~\cite{kaplan2021} and therefore should not alter our results drastically. 

Another interesting effect is the impact of valley polarization on the shift current and the photoresponse in general. Our shift-current expression considered in Eq.~\ref{shiftvector} has equal contributions from both valleys if the Dirac cones of the underlying graphene layers are not tilted. However, in addition to the shift-current contribution which comes with a Dirac-delta function, the second-order conductivity also has a contribution from the principal part as presented in Eq.~\ref{s23box} of Appendix~\ref{Appendix:shift_current}. This contribution is equal and opposite from two valleys and hence can affect the shift-current response for a valley-polarized setup only. This valley dependence would be even more apparent for injection currents.

As pointed out earlier, the shift currents are a reflection of the quantum geometry of the electronic bands. The effect is also clearly related to the charge distribution of the Bloch wave functions in the gigantic Moir\'e unit cell. Indeed, as argued in the Sec. \ref{interactionsMF}, different momentum states lead to a different spatial distribution of charge, e.g. for flat-bands $\kappa$ points states give rise to charge buildup near $AA$ sites whilst $\gamma$ point states cause a buildup of charge in a ring surrounding $AA$ sites. For the first dispersive bands however, the relation flips - $\kappa$ points states give rise to charge buildup in a ring surrounding $AA$ sites whilst $\gamma$ point states lead to a charge buildup at the $AA$ sites. We find that qualitatively sharp resonances seen in Fig.\ref{FDHartree} correspond precisely to the transitions for $AA$ charge profile to that of a ring surrounding the $AA$ sites or vice versa. In future work, it would be fascinating to consider what additional effects emerge from these unusual rearrangements of the electronic probability density within the moir\'e unit cell.

\section{Acknowledgment}
We thank Stevan Nadj-Perge for an earlier collaboration and useful discussions. We  acknowledge  support  from  the  Institute  of  Quantum Information  and  Matter,  an  NSF  Physics  Frontiers  Center funded  by  the  Gordon  and  Betty  Moore  Foundation,  the Packard  Foundation,  and  the  Simons  Foundation.    G.R.  and S.C are  grateful  for  support  from  the U.S. Department of Energy, Office of Science, Basic Energy Sciences under  Award  desc0019166. GR  is  also  grateful  to  the NSF DMR grant number 1839271. C.L. acknowledges support from the Gordon and Betty Moore Foundation through Grant GBMF8682.

\onecolumngrid

 \twocolumngrid

\clearpage
\bibliography{tblg.bib}

\clearpage
\begin{widetext}
\appendix

\renewcommand{\thefigure}{S\arabic{figure}}
\renewcommand{\figurename}{Supplemental Figure}
\setcounter{figure}{0}
\begin{center}
    \bf{Supplementary material for ``Shift-current response as a probe of quantum geometry and electron-electron interactions in twisted bilayer graphene''}
\end{center}

 \begin{figure}[h]
 	\includegraphics[scale=0.42]{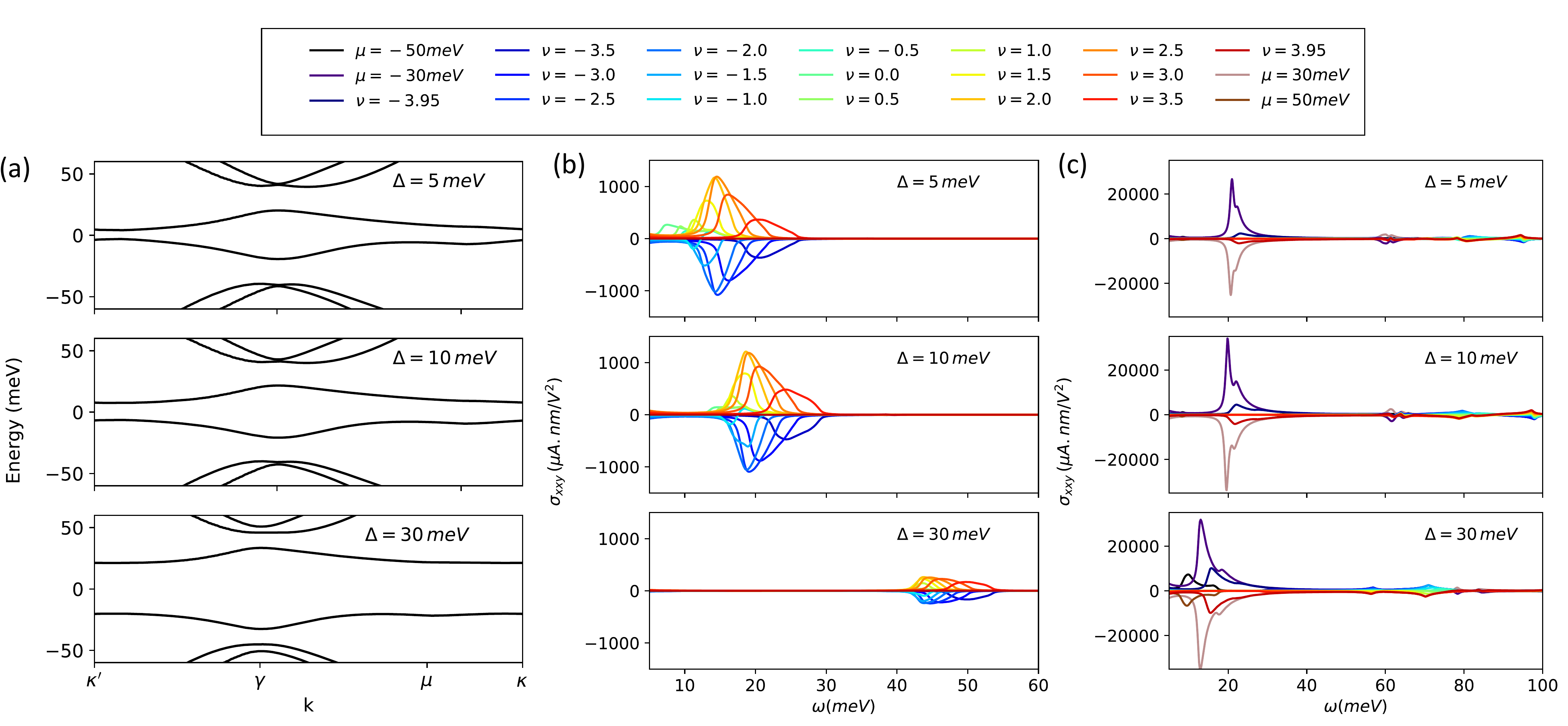}
 	\caption{\textbf{Effect of increasing sublattice offset energy, $\Delta$ on energy spectrum and shift-current response :} (a) Energy spectrum around flat-bands, (b) FF contribution to seconf-order conductivity, and (c) FD contribution to second-order conductivity of twisted bilayer graphene with twist angle, $\theta=0.8^\circ$, for three different sublattice offset energies. As $\Delta$ inceases, the gap between flat bands increases but they come closer to the dispersive bands and it results in an opposite frequency shift for the peak value in FF and FD case.}
 	\label{Fig_Deltachange}
 \end{figure}   
 
\section{Group Theoretical Analysis for second-order conductivity}
\label{Appendix:Symm}
If we consider a TBG  encapsulated with hBN from both sides such that the sublattice symmetry breaking effect is same on both layers, we have $\Delta_1=\Delta_2$. In this case, the symmetry group of TBG is $D_3$ generated by $C_{3z}$ and $C_{2y}$. In our simulations, we found that there is only one independent component of $\sigma_{\alpha\beta}^{\mu}$ tensor when $\Delta_1=\Delta_2$. It can be directly deduced from the number of cubic functions associated with the trivial irrep $A_1$ in the character table of $D_3$ shown in Tab.~\ref{tableD3}. 

\begin{table}[h]
	\begin{tabular}{|c|c|c|c|c|c|c|}
		\hline
		Irreps&$E$&$2C_{3z}$&$3C_{2y}$& Linear functions &Quadratic functions& Cubic functions\\
		\hline
		$A_1$ &1& 1 & 1& - &$x^2+y^2$, $z^2$&$y(y^2-3x^2)$\\
		\hline 
		$A_2$ & 1 & 1 & -1& $z,R_z$ &- & $z^3, x(3y^2-x^2), z(x^2+y^2)$\\
		\hline
		$E$ & 2 & -1 & 0 & $(x,y),(R_x,R_y)$ & $(y^2-x^2,xy)(xz,yz)$ & $(xz^2,yz^2)$ $\left[xyz,z(y^2-x^2)\right]$ $\left[y(x^2+y^2),x(x^2+y^2)\right]$\\
		\hline
	\end{tabular}
	\caption{Character table for point group $D_3$ generated by $C_{3z}$ and $C_{2y}$.}

\label{tableD3}
\end{table}
\begin{table}[h]
	\begin{tabular}{|c|c|c|c|c|c|c|}
		\hline
		Irreps&$e$&$c$&$c^2$& Linear functions &Quadratic functions& Cubic functions\\
		\hline
		$A_1$ &1& 1 & 1& $z$ &$x^2+y^2$, $z^2$&$z^3$,$y(y^2-3x^2)$,$y(y^2-3x^2)$,$z(x^2+y^2)$\\
		\hline \begin{tabular}{c}$E$\end{tabular}
		& \begin{tabular}{c}1\\1\end{tabular}   &  \begin{tabular}{c}$e^{i\frac{2\pi}{3}}$\\$e^{-i\frac{2\pi}{3}}$\end{tabular} & \begin{tabular}{c}$e^{-i\frac{2\pi}{3}}$\\$e^{i\frac{2\pi}{3}}$\end{tabular} & \begin{tabular}{c}$x+iy, R_x+iR_y$\\$x-iy, R_x-iR_y$\end{tabular}  &\begin{tabular}{c}$\left(x^2-y^2,xy\right)$ $\left(yz,xz\right)$\end{tabular} & $(xz^2,yz^2)$ $ \left[xyz,z(x^2-y^2)\right]$ $\left[x(x^2+y^2),y(x^2+y^2)\right]$\\

		\hline
	\end{tabular}
	\caption{Character table for point group $C_{3z}$.}
	
	\label{tableC3}
\end{table}

It indicates that the second order tensor $\sigma_{\alpha\beta}^{\mu}$ has only one independent element with
\begin{equation}
\begin{split}
\sigma_{xx}^y=-\sigma_{yy}^y=\sigma_{xy}^y=\sigma_{yx}^x\ne 0\\
\sigma_{yy}^x=-\sigma_{xx}^x=\sigma_{yx}^y=\sigma_{xy}^y=0.
\end{split}
\end{equation} 
On the other hand, as shown in Fig.~\ref{tableC3}, the trivial irrep for $C_{3z}$ has two cubic functions (ignoring the ones involving $z$ as our system is two dimensional only) indicating that a rank three tensor can have two independent components under $C_3$. As a result of this, for $\Delta_1\ne \Delta_2$, we have
\begin{equation}
\begin{split}
\sigma_{xx}^y=-\sigma_{yy}^y=\sigma_{xy}^y=\sigma_{yx}^x\ne 0\\
\sigma_{yy}^x=-\sigma_{xx}^x=\sigma_{yx}^y=\sigma_{xy}^y\ne 0
\end{split}
\end{equation} 
which is consistent with our observation in Fig.~\ref{fig:differentdelta}.
\begin{figure}
    \centering
    \includegraphics[scale=0.4]{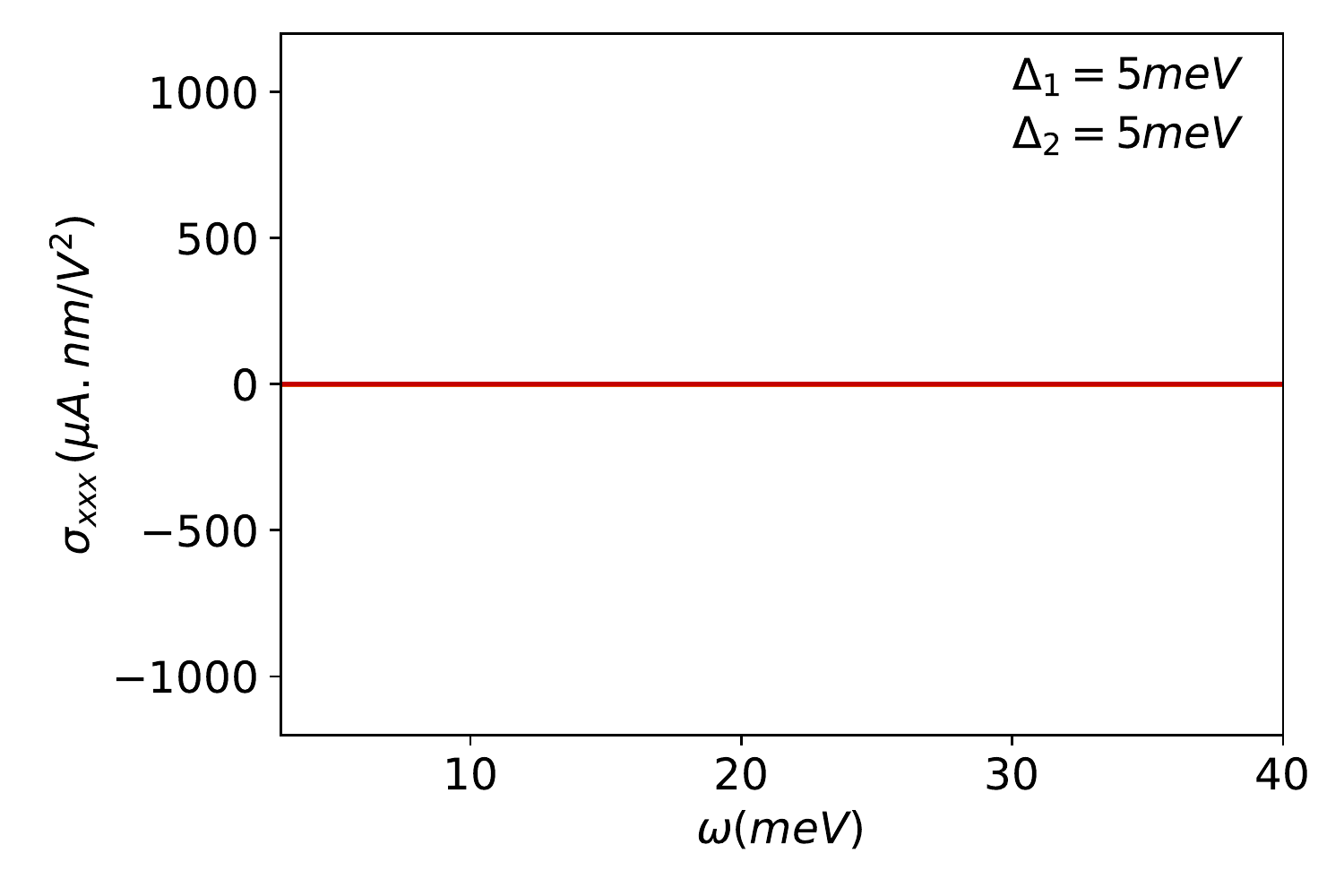}
    \includegraphics[scale=0.4]{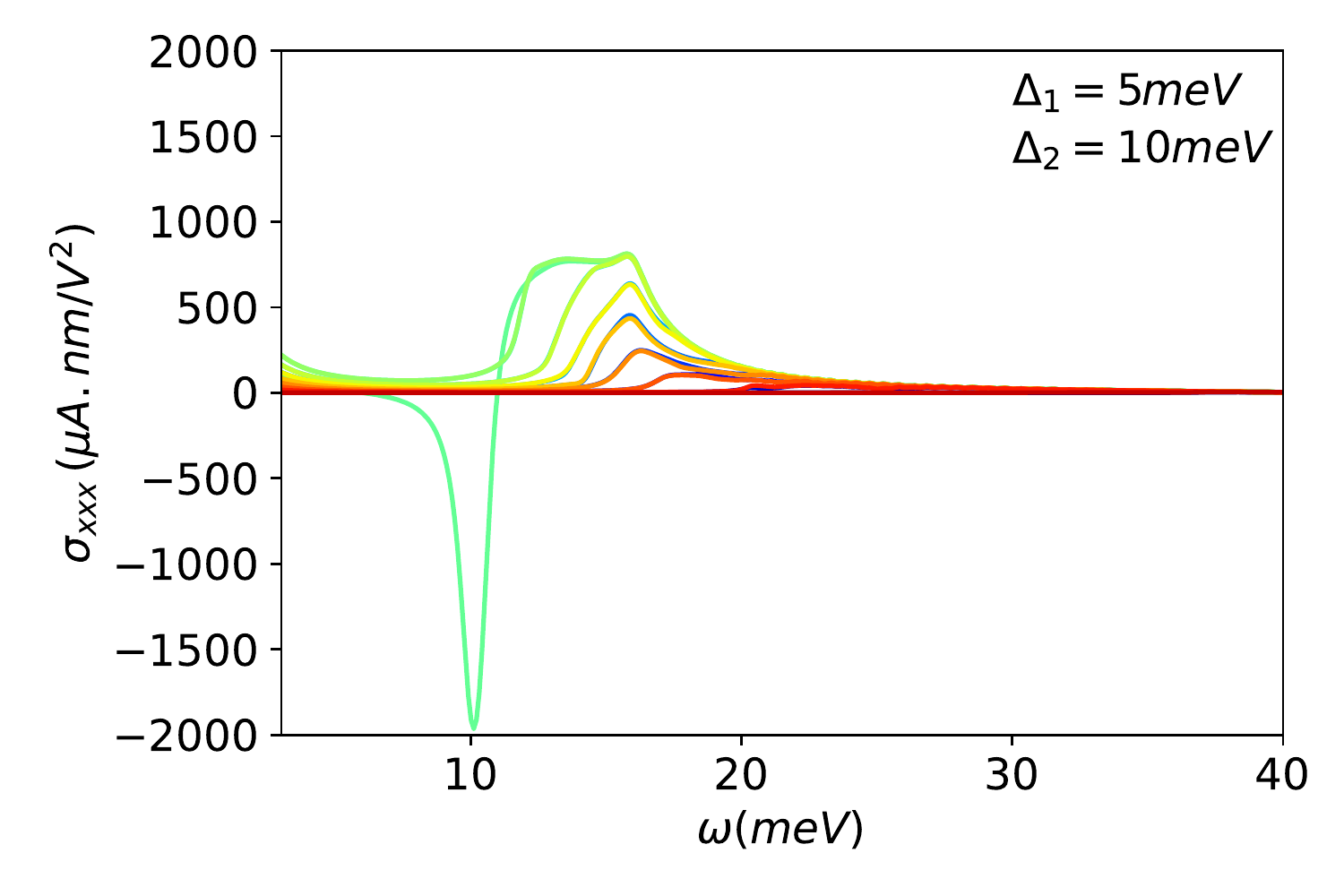}
    \includegraphics[scale=0.4]{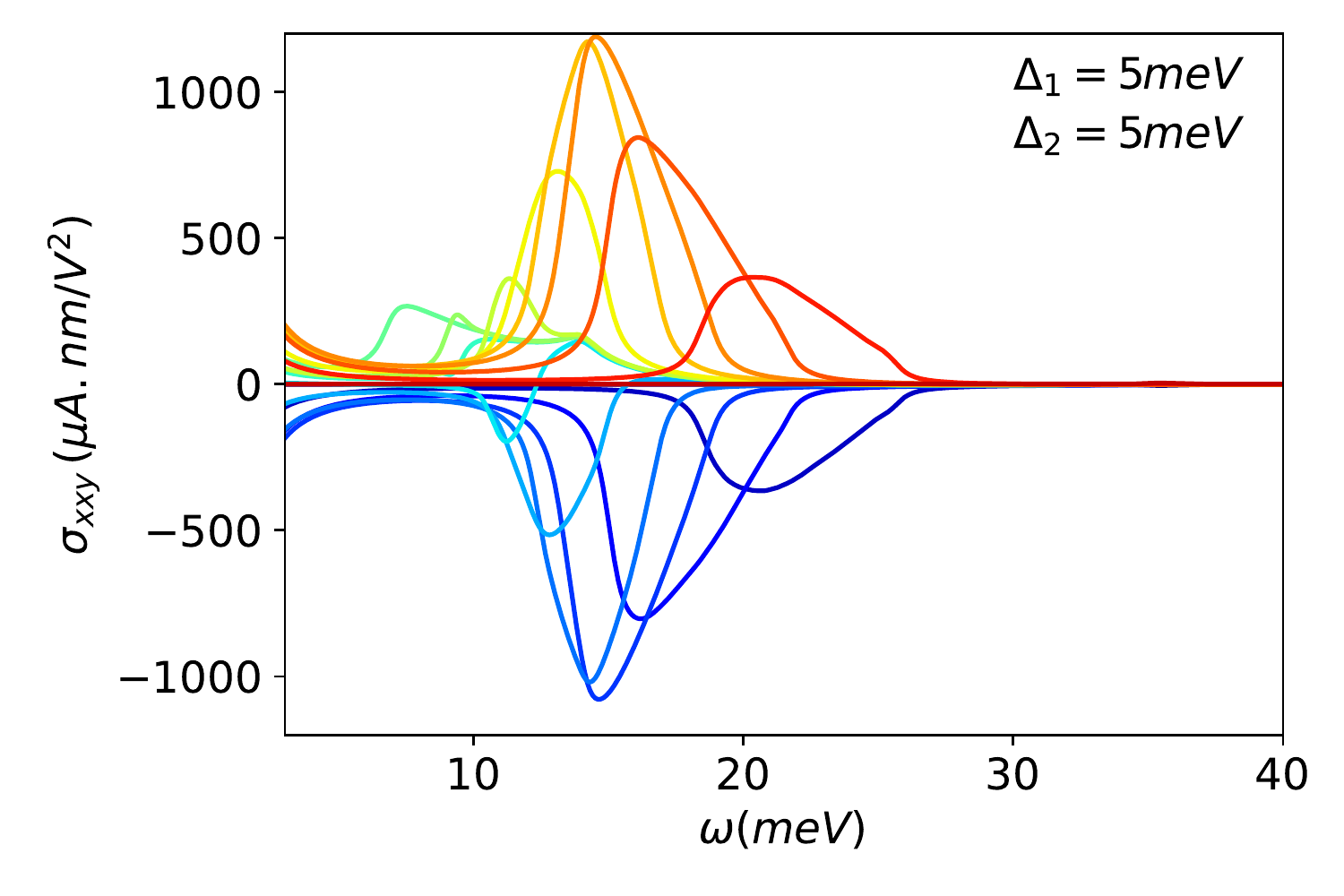}
    \includegraphics[scale=0.4]{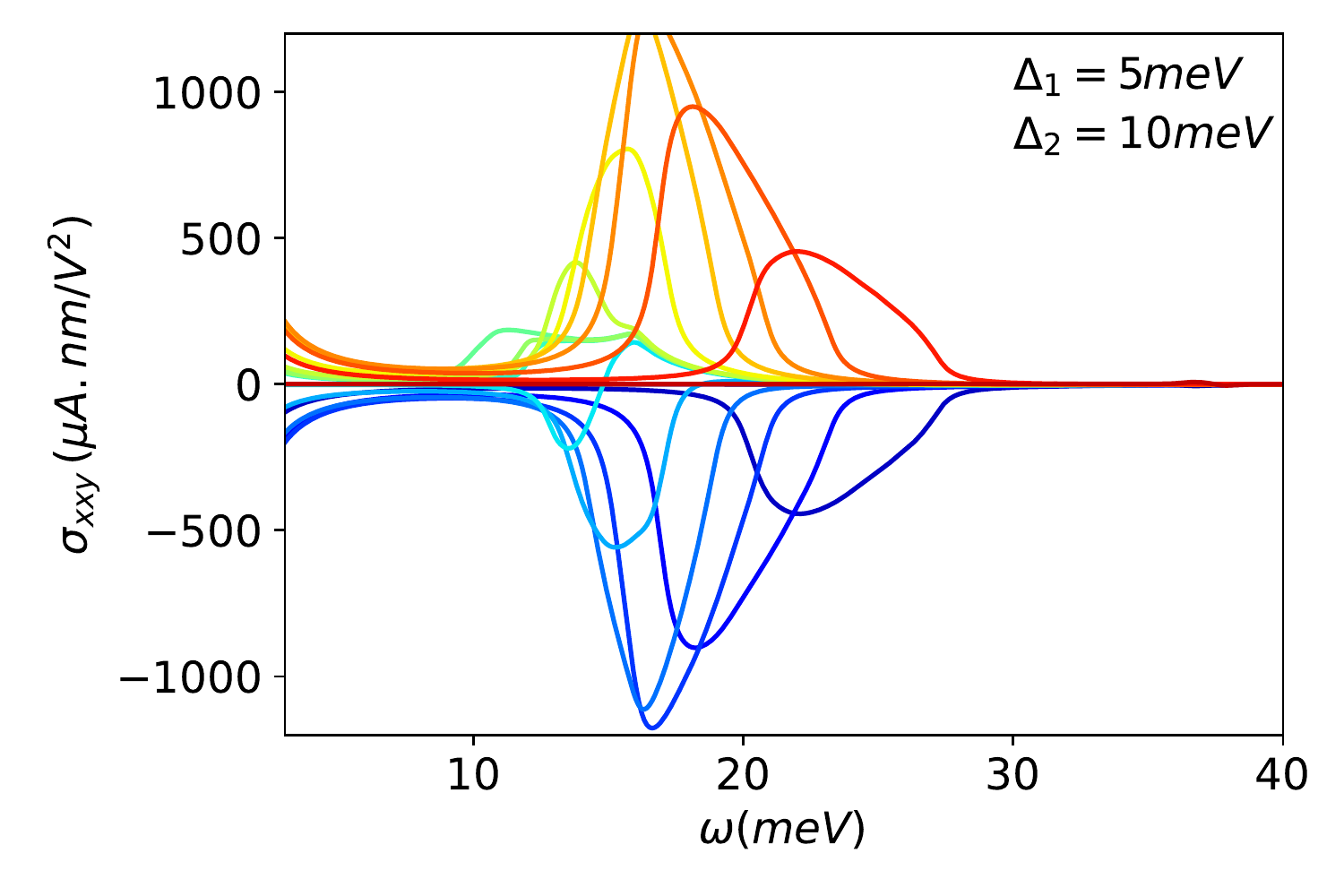}
    \includegraphics[scale=0.4]{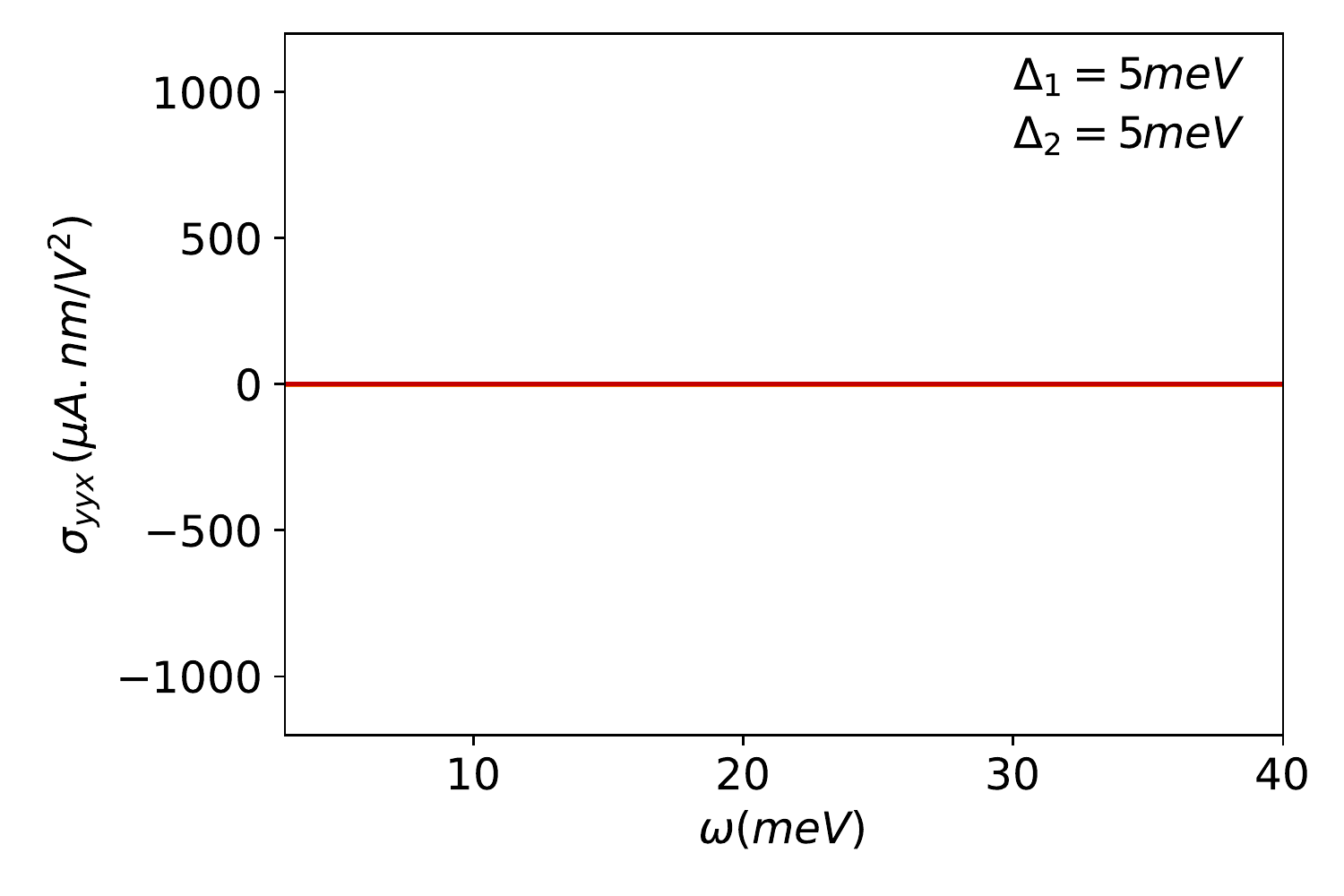}
    \includegraphics[scale=0.4]{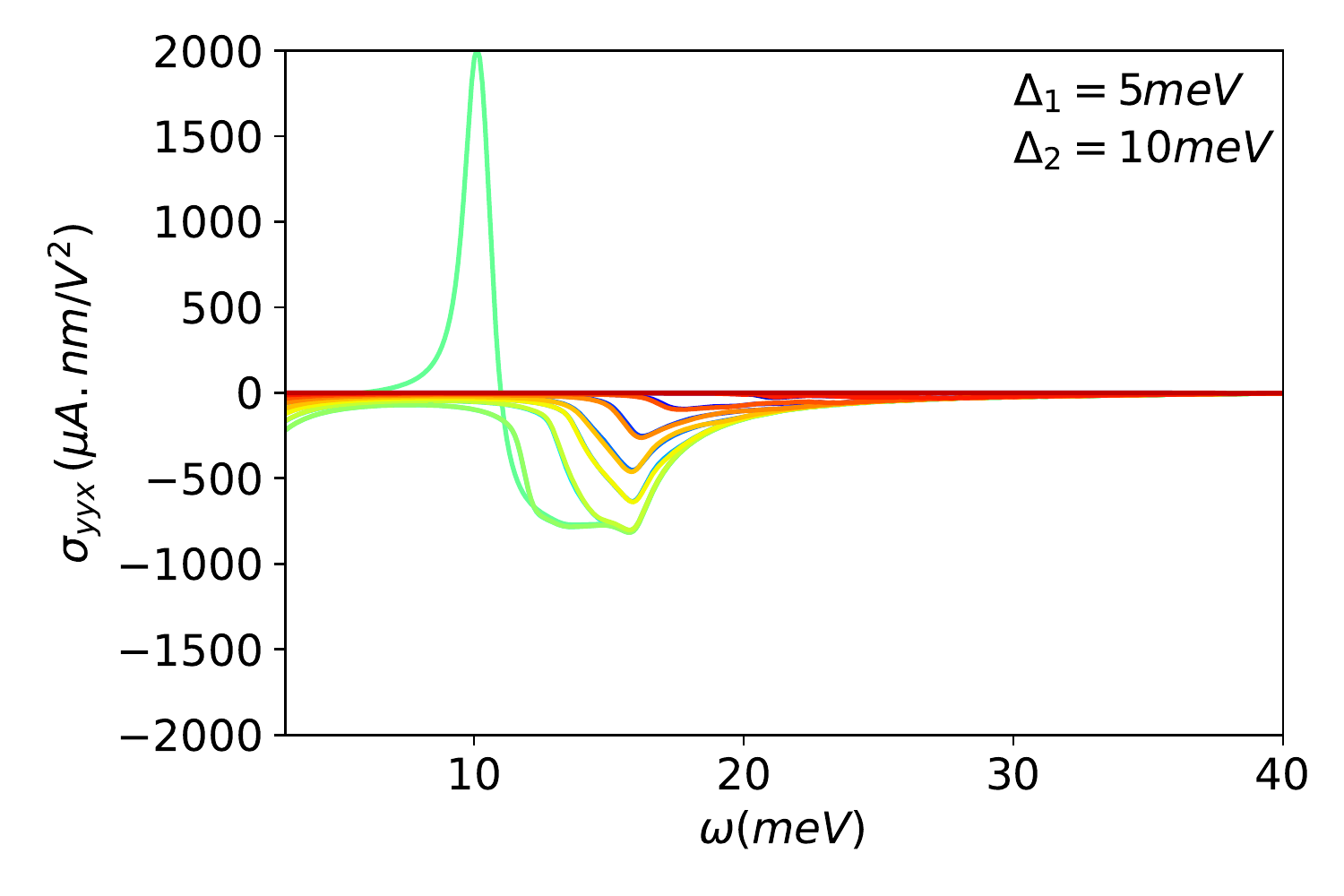}
    \includegraphics[scale=0.4]{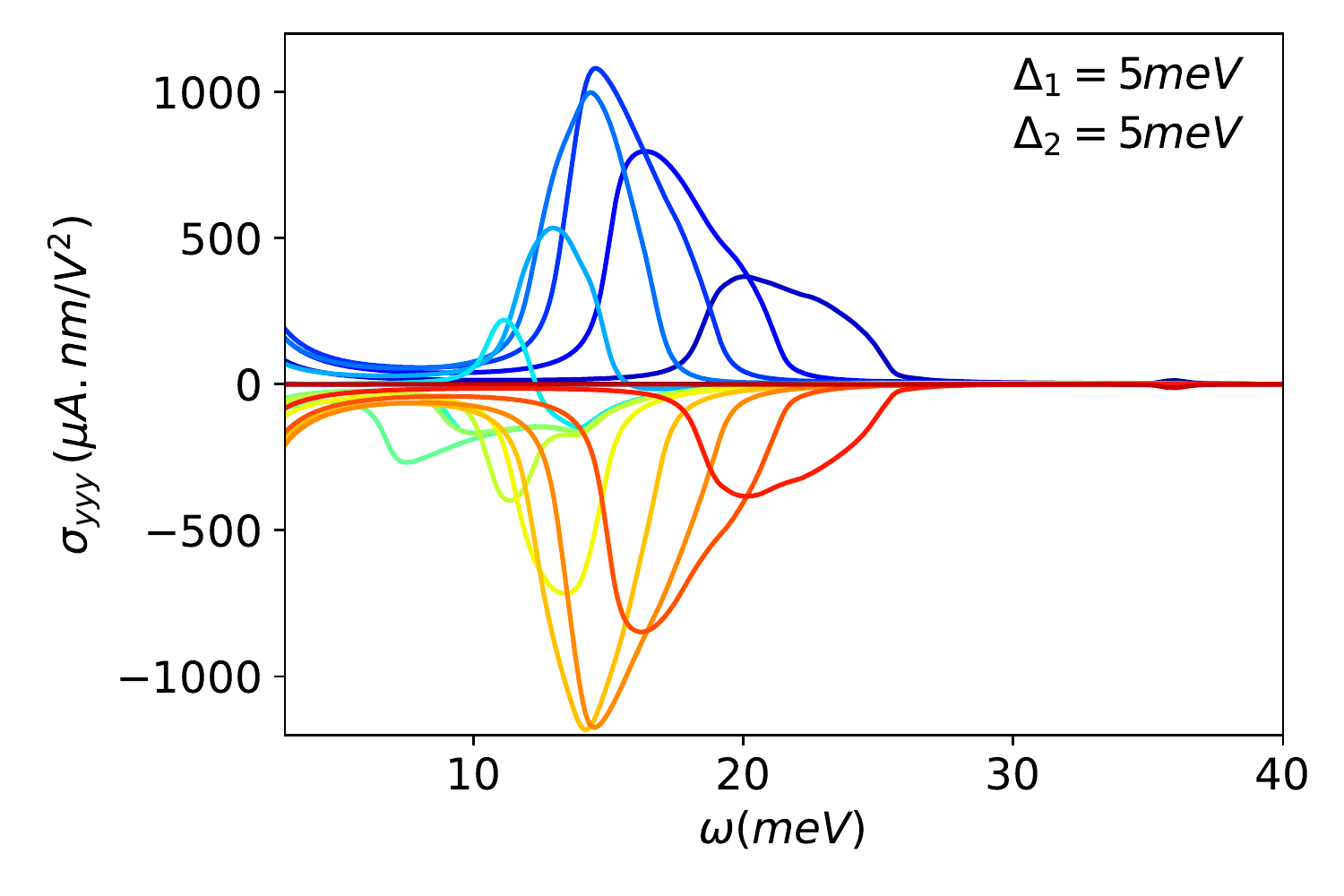}
    \includegraphics[scale=0.4]{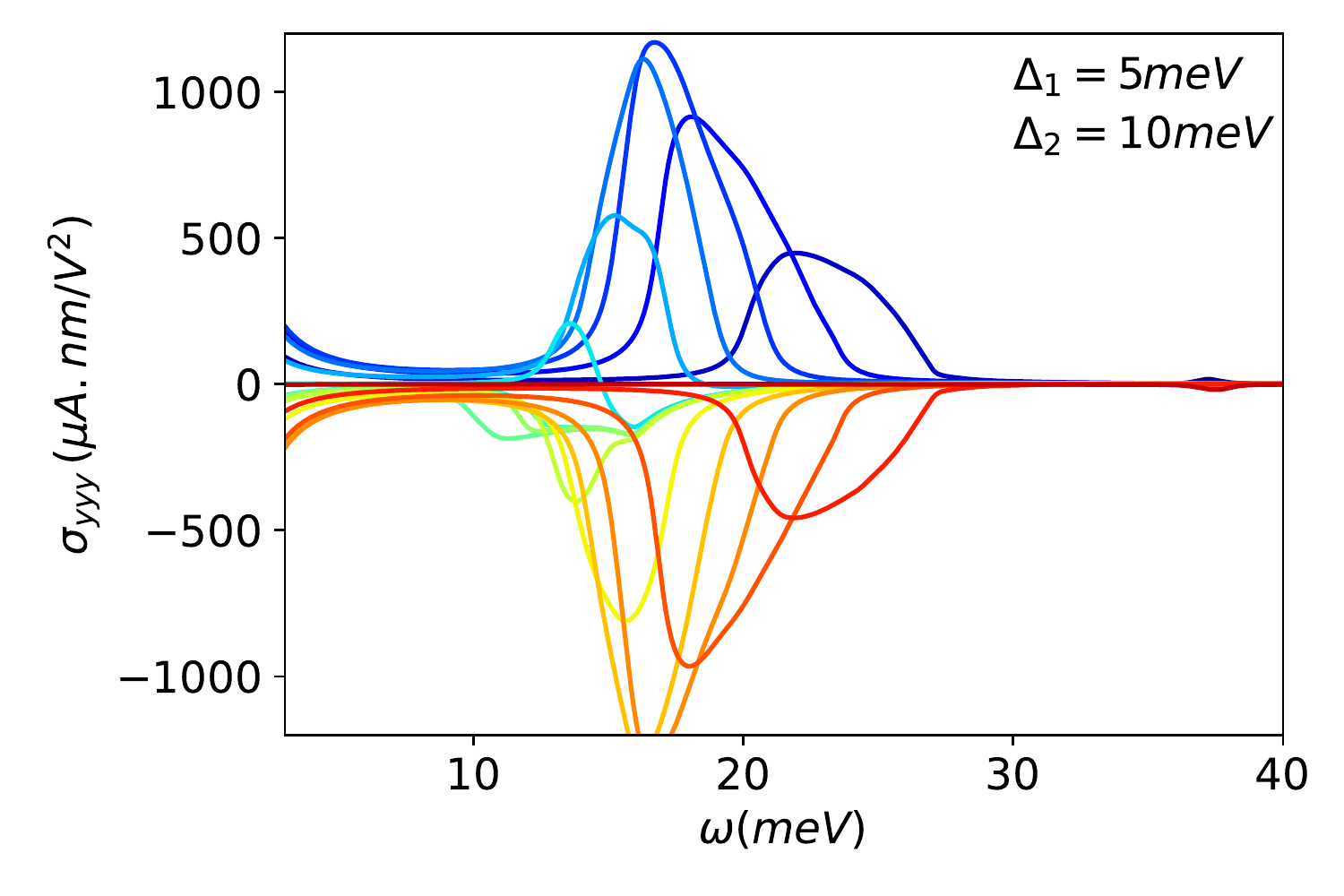}
    \caption{\textbf{Effect of $C_{2y}$ symmetry breaking on different shift-current response.} Second-order conductivity tensor elements, $\sigma_{xxx}$,$\sigma_{xxy}$, $\sigma_{yyx}$, and $\sigma_{yyy}$ for $\Delta_1=\Delta_2=\Delta$ case (left column) and for $\Delta_1\ne\Delta_2$ (right column). We notice that breaking $C_{2y}$  by choosing different values for $\Delta_1$ and $\Delta_2$, results in a non-zero value for $\sigma_{xxx}$ and $\sigma_{yyy}$.}
    \label{fig:differentdelta}
\end{figure}

\subsection*{Helicity dependent current in TBG}
\label{note2}
The second-order CPGE is captured by a rank two tensor (not a rank three tensor like shift-current response)
\begin{equation}j_\alpha=i\eta_{\alpha\beta}(\mathbf{E}\times\mathbf{E}^*)_\beta.\end{equation} 
If we consider a two-dimensional material in $x-y$ plane, then the normal incidence results in $\beta=z$ and the in-plane current requires, $\alpha=x,y$. Now, the quantity $(\mathbf{E}\times\mathbf{E}^*)_z$ is the $z$ component of an axial vector which shares representation $A_1$ in $C_{3z}$ character table~\ref{tableC3} (denoted by $R_z$ in the character table). On the other hand, the electric current, $\mathbf{j}=j_x\hat{x}+j_y\hat{y}$ is a polar vector which has irrep $E$, and thus the CPGE conductivity tensor, $n_{\alpha\beta}$ transforms according to irrep $A_1\otimes E=E$ which contains no trivial irrep $A_1$, and thus $n_{\alpha z}=0$. However, if we consider an oblique incidence (i.e $\beta=x,y$), then we might get a non-zero component but this needs an electric field with $z$ component. Now, in the minimal coupling picture, we do not have a momentum component in this direction and $E_z$ cannot couple to our system and thus CPGE can not occur. However, it can couple to the layer degree of freedom which would be an interesting direction to pursue but we need to go beyond the minimal coupling approach which is beyond the scope of this current work. However, if the symmetry is lowered to a reflection symmetry with a mirror plane perpendicular to the plane of the bilayer graphene, it can result in a non-zero injection current from circularly polarized light. 
\section{Shift current expressions}
\label{Appendix:shift_current}
Within the independent particle approximation and using minimal coupling approach, the second-order conductivity for a perturbation arising from a linearly polarized EM field can be obtained using formula Eq. 43 of Ref.\cite{Parker19}
\begin{equation}
\begin{split}
\sigma^{\mu}_{\alpha\beta}(\omega_Z,\omega_1,\omega_2)=&-\frac{e^3}{\hbar^2 \omega_1 \omega_2}\sum_{a,b,c}\int d\textbf{k} \,\frac{1}{2}f_a h_{aa}^{\mu\alpha\beta}+\frac{1}{2}f_a h_{aa}^{\mu\beta\alpha}+f_{ab}\frac{h_{ab}^\alpha h_{ba}^{\mu\beta}}{\omega_1-\epsilon_{ab}}+f_{ab}\frac{h_{ab}^\beta h_{ba}^{\mu\alpha}}{\omega_2-\epsilon_{ab}}+\frac{1}{2}f_{ab}\frac{h_{ab}^{\alpha\beta} h_{ba}^{\mu}}{\omega_Z-\epsilon_{ab}}\\&+\frac{1}{2}f_{ab}\frac{h_{ab}^{\beta\alpha} h_{ba}^{\mu}}{\omega_Z-\epsilon_{ab}}+\frac{h_{ab}^{\alpha}h_{bc}^\beta h_{ca}^\mu}{\omega_Z-\epsilon_{ca}}\left[\frac{f_{ab}}{\omega_1-\epsilon_{ba}}+\frac{f_{cb}}{\omega_2-\epsilon_{cb}}\right]+\frac{h_{ab}^{\beta}h_{bc}^\alpha h_{ca}^\mu}{\omega_Z-\epsilon_{ca}}\left[\frac{f_{ab}}{\omega_2-\epsilon_{ba}}+\frac{f_{cb}}{\omega_1-\epsilon_{cb}}\right]
\end{split}
\end{equation}
where $\omega_Z=\omega_1+\omega_2$,  $h^{\alpha}_{ab}=\left<a|\nabla_{k_\alpha}H|b\right>$, $h^{\alpha\beta}_{ab}=\left<a|\nabla_{k_\alpha}\nabla_{k_\beta}H|b\right>$ are derivatives of hamiltonian, $\epsilon_{ab}=\epsilon_{a}-\epsilon_{b}$ is the energy difference, and $f_{ab}=f_a-f_b$ is the difference in occupancy of energy level $a$ and $b$. This formula many different contributions like injection current, shift current etc. It can be recast in a slightly different form by shifting 
 all frequencies by $\omega\rightarrow\omega+i\eta$ and the above equation reduces to 
\begin{equation}
\begin{split}
\sigma^{\mu}_{\alpha\beta}(\omega_Z,\omega_1,\omega_2)=-\frac{e^3}{\hbar^2 \omega_1 \omega_2}\sum_{a,b,c}\int d\textbf{k} \,\frac{1}{2}f_a h_{aa}^{\mu\alpha\beta}+\frac{1}{2}f_a h_{aa}^{\mu\beta\alpha}+f_{ab}\frac{h_{ab}^\alpha h_{ba}^{\mu\beta}}{\omega_1+i\eta-\epsilon_{ab}}+f_{ab}\frac{h_{ab}^\beta h_{ba}^{\mu\alpha}}{\omega_2+i\eta-\epsilon_{ab}}+\frac{1}{2}f_{ab}\frac{h_{ab}^{\alpha\beta} h_{ba}^{\mu}}{\omega_Z+i\eta-\epsilon_{ab}}&\\+\frac{1}{2}f_{ab}\frac{h_{ab}^{\beta\alpha} h_{ba}^{\mu}}{\omega_Z+i\eta-\epsilon_{ab}}+\frac{h_{ab}^{\alpha}h_{bc}^\beta h_{ca}^\mu}{\omega_Z+i\eta-\epsilon_{ca}}\left[\frac{f_{ab}}{\omega_1+i\eta-\epsilon_{ba}}+\frac{f_{cb}}{\omega_2+i\eta-\epsilon_{cb}}\right]+\frac{h_{ab}^{\beta}h_{bc}^\alpha h_{ca}^\mu}{\omega_Z+i\eta-\epsilon_{ca}}\left[\frac{f_{ab}}{\omega_2+i\eta-\epsilon_{ba}}+\frac{f_{cb}}{\omega_1+i\eta-\epsilon_{cb}}\right]
\end{split}
\label{Eq.43full}
\end{equation}
The DC response to an AC field of frequency $\omega$ is given by $\sigma^\mu_{\alpha\beta}(0,\omega,-\omega)$. 
 which can be obtained from Eq.~\ref{Eq.43full} by substituting $\omega_1=-\omega_2=\omega$ and for our prime case of interest ($\alpha=\beta$), we get
\begin{equation}
\begin{split}
\sigma^{\mu}_{\alpha\alpha}(0,\omega,-\omega)=\frac{e^3}{\hbar^2 \omega^2}\sum_{a,b,c}\int d\textbf{k} \,f_a h_{aa}^{\mu\alpha\alpha}+f_{ab}\frac{h_{ab}^\alpha h_{ba}^{\mu\alpha}}{\omega+i\eta-\epsilon_{ab}}+f_{ab}\frac{h_{ab}^\alpha h_{ba}^{\mu\alpha}}{-\omega+i\eta-\epsilon_{ab}}+f_{ab}\frac{h_{ab}^{\alpha\alpha} h_{ba}^{\mu}}{\epsilon_{ba}}&\\+\frac{h_{ab}^{\alpha}h_{bc}^\alpha h_{ca}^\mu}{\epsilon_{ac}}\left[\frac{f_{ab}}{\omega+i\eta-\epsilon_{ba}}+\frac{f_{cb}}{-\omega+i\eta-\epsilon_{cb}}\right]+\frac{h_{ab}^{\alpha}h_{bc}^\alpha h_{ca}^\mu}{\epsilon_{ac}}\left[\frac{f_{ab}}{-\omega+i\eta-\epsilon_{ba}}+\frac{f_{cb}}{\omega+i\eta-\epsilon_{cb}}\right]
\end{split}
%\label{shiftcurrent1}
\end{equation}
\subsection{Connections with shift current expression}
In order to understand the connections between the shift-current expression we encountered in the main text and the form the second-order conductivity considered above, we can first split this equation into two different kind of contributions
\begin{equation}
\begin{split}
\sigma^{\mu}_{\alpha\alpha}(0,\omega,-\omega)=\frac{e^2}{\hbar^2 \omega^2}\sum_{a,b,c}\int d\textbf{k} \,f_a h_{aa}^{\mu\alpha\alpha}+\underbrace{f_{ab}\frac{h_{ab}^\alpha h_{ba}^{\mu\alpha}}{\omega+i\eta-\epsilon_{ab}}+f_{ab}\frac{h_{ab}^\alpha h_{ba}^{\mu\alpha}}{-\omega+i\eta-\epsilon_{ab}}}_{\sigma^{23}}+f_{ab}\frac{h_{ab}^{\alpha\alpha} h_{ba}^{\mu}}{\epsilon_{ba}}+\\\underbrace{\frac{h_{ab}^{\alpha}h_{bc}^\alpha h_{ca}^\mu}{\epsilon_{ac}}\left[\frac{f_{ab}}{\omega+i\eta-\epsilon_{ba}}+\frac{f_{cb}}{-\omega+i\eta-\epsilon_{cb}}\right]+\frac{h_{ab}^{\alpha}h_{bc}^\alpha h_{ca}^\mu}{\epsilon_{ac}}\left[\frac{f_{ab}}{-\omega+i\eta-\epsilon_{ba}}+\frac{f_{cb}}{\omega+i\eta-\epsilon_{cb}}\right]}_{\sigma^{56}}
\end{split}
\label{shiftcurrent1}
\end{equation}
Let's first focus on 2nd and 3rd term of Eq.~\ref{shiftcurrent1}, $\sigma^{23}$,  where the integrand  can be expressed as 
\begin{equation}
\begin{split}
f_{ab}\frac{h_{ab}^\alpha h_{ba}^{\mu\alpha}}{\omega+i\eta-\epsilon_{ab}}+f_{ab}\frac{h_{ab}^\alpha h_{ba}^{\mu\alpha}}{-\omega+i\eta-\epsilon_{ab}}=f_{ab}\frac{h_{ab}^\alpha h_{ba}^{\mu\alpha}}{\omega+i\eta-\epsilon_{ab}}-f_{ba}\frac{h_{ba}^\alpha h_{ab}^{\mu\alpha}}{\omega-i\eta+\epsilon_{ba}}=f_{ab}\frac{h_{ab}^\alpha h_{ba}^{\mu\alpha}}{\omega+i\eta-\epsilon_{ab}}+\\f_{ab}\frac{h_{ba}^\alpha h_{ab}^{\mu\alpha}}{\omega-i\eta-\epsilon_{ab}}=f_{ab}P\left(\frac{1}{\omega-\epsilon_{ab}}\right)\left[h_{ab}^\alpha h_{ba}^{\mu\alpha}+h_{ba}^\alpha h_{ab}^{\mu\alpha}\right]+f_{ab}i\pi \left[h_{ab}^\alpha h_{ba}^{\mu\alpha}-h_{ba}^\alpha h_{ab}^{\mu\alpha}\right]\delta(\omega-\epsilon_{ab})
\end{split}
\label{sndthrd}
\end{equation}
For our purpose, the most interesting term is the one involving $\delta(\omega-\epsilon_{ab})$. We can write 
\begin{equation}
h_{ab}^\alpha h_{ba}^{\mu\alpha}-h_{ba}^\alpha h_{ab}^{\mu\alpha}=h_{ab}^\alpha h_{ba}^{\mu\alpha}-[a\leftrightarrow b ].
\label{abexp}
\end{equation}
Now, first we derive an expression for $h_{mn}^{\alpha\beta}$. According to the notation used in Ref.~\cite{Parker19},
\begin{equation}
h_{mn}^{\alpha\beta}=\left[D^\alpha D^\beta[H_0]\right]_{mn}\equiv \left<m|\nabla_\alpha\nabla_\beta(H_0)|n\right>
\end{equation}
where $H_0$ is the unperturbed hamiltonian and for a given operator $O$
\begin{equation}
D[O]_{ab}=[D,O]_{ab}=\nabla_k(O_{ab})-i[\mathbf{A},O]_{ab}
\end{equation}
where $\mathbf{A}$ is the Berry-connection matrix with $\mathbf{A}^\mu_{mn}=i\left<u_m|\partial_{k^\mu}|u_n\right>$. We can thus write
\begin{equation}
h_{mn}^{\alpha\beta}=\left[D^\alpha D^\beta[H_0]\right]_{mn}=\partial_\alpha\left(\left[D^\beta[H_0]\right]_{mn}\right)-i\left[\mathbf{A}^\alpha, D^\beta[H_0]\right]_{mn}.
\label{ddalpha}
\end{equation}
We have
\begin{equation}
 h^\beta_{mn}=\left(D^\beta[H_0]\right)_{mn}=\partial_\beta((H_0)_{mn})-i\left[\mathbf{A}^\beta, H_0\right]_{mn}=\underbrace{\delta_{mn}v^\beta_{nn}-i(\epsilon_n-\epsilon_m)\mathbf{A}^\beta_{mn}}_ {v^\beta_{mn}}
 \label{vmn}
\end{equation}
where we have used the fact that $\left<m|H_0|n\right>=\delta_{mn}\epsilon_{n}$ and $v^\beta_{nn}=\partial_\beta \epsilon_n$. We can express
\begin{equation}
\left[\mathbf{A}^\alpha, D^\beta[H_0]\right]_{mn}=\left[\mathbf{A}^\alpha h^\beta-h^\beta \mathbf{A}^\alpha\right]_{mn}=\mathbf{A}^\alpha_{md}h^\beta_{dn}-\mathbf{A}^\alpha_{dn}h^\beta_{md}
\label{comm1}
\end{equation}
For the first term in Eq.~\ref{ddalpha}, we can use Eq.~\ref{vmn} to write
\begin{equation}
\partial_\alpha\left(\left[D^\beta[H_0]\right]_{mn}\right)=\delta_{mn}\partial_\alpha\epsilon_{n}-i(v_{nn}^\alpha-v_{mm}^\alpha)\mathbf{A}^\beta_{mn}-i\epsilon_{nm}\partial_\alpha\mathbf{A}^\beta_{mn}
\end{equation}
and the second part can be fully extended using Eq.~\ref{vmn} and Eq.~\ref{comm1}
\begin{equation}
-i\left[\mathbf{A}^\alpha, D^\beta[H_0]\right]_{mn}=-i\mathbf{A}^\alpha_{md}\left(\delta_{dn}v^\beta_{nn}-i\epsilon_{nd}\mathbf{A}^\beta_{db}\right)+i\mathbf{A}^\alpha_{dn}\left(\delta_{md}v^\beta_{mm}-i\epsilon_{dm}\mathbf{A}^\beta_{md}\right)
\end{equation}
Now, combining these two equations we get:
\begin{equation}
\begin{split}
h_{mn}^{\alpha\beta}=\partial_\alpha\left(\left[D^\beta[H_0]\right]_{mn}\right)-i\left[\mathbf{A}^\alpha, D^\beta[H_0]\right]_{mn}=\delta_{mn}\partial_\alpha\epsilon_{n}-i(v_{nn}^\alpha-v_{mm}^\alpha)\mathbf{A}^\beta_{mn}-i\epsilon_{nm}\partial_\alpha\mathbf{A}^\beta_{mn}\\-i\mathbf{A}^\alpha_{md}\delta_{dn}v^\beta_{nn}+i\mathbf{A}^\alpha_{dn}\delta_{md}v^\beta_{mm}-\epsilon_{nd}\mathbf{A}^\alpha_{md}\mathbf{A}^\beta_{dn}+\epsilon_{dm}\mathbf{A}^\alpha_{dn}\mathbf{A}^\beta_{md}.
\end{split}
\end{equation}
\begin{equation}
\begin{split}
h_{mn}^{\alpha\beta}=\partial_\alpha\left(\left[D^\beta[H_0]\right]_{mn}\right)-i\left[\mathbf{A}^\alpha, D^\beta[H_0]\right]_{mn}=\delta_{mn}\partial_\alpha\epsilon_{n}-i(v_{nn}^\alpha-v_{mm}^\alpha)\mathbf{A}^\beta_{mn}-i\epsilon_{nm}\partial_\alpha\mathbf{A}^\beta_{mn}\\-i\mathbf{A}^\alpha_{mn}v^\beta_{nn}+i\mathbf{A}^\alpha_{mn}v^\beta_{mm}-\epsilon_{nd}\mathbf{A}^\alpha_{md}\mathbf{A}^\beta_{dn}+\epsilon_{dm}\mathbf{A}^\alpha_{dn}\mathbf{A}^\beta_{md}.
\end{split}
\end{equation}
Our goal was to evaluate $h_{ab}^\alpha h_{ba}^{\mu\alpha}$ in  Eq.~\ref{abexp}. For now, we are going to focus on case $a\ne b$
For $a\ne b$, $h_{ab}^\alpha=-i\epsilon_{ba}\mathbf{A}^\alpha_{ab}$ from Eq.~\ref{vmn} and similarly 
\begin{equation}
h_{ab}^\alpha h_{ba}^{\mu\alpha}=-i\epsilon_{ba}\mathbf{A}^\alpha_{ab}\left(-i\Delta_{ab}^\mu\mathbf{A}^\alpha_{ba}-i\epsilon_{ab}\partial_\mu\mathbf{A}_{ba}^\alpha-i\mathbf{A}^\mu_{ba}\Delta_{ab}^\alpha-\epsilon_{ad}\mathbf{A}^\mu_{bd}\mathbf{A}^\alpha_{da}+\epsilon_{db}\mathbf{A}^\mu_{da}\mathbf{A}^\alpha_{bd}\right).
\end{equation}
where $\Delta_{ab}^\mu=v^\mu_{aa}-v^\mu_{bb}$.
This gives
\begin{equation}
\begin{split}
h_{ab}^\alpha h_{ba}^{\mu\alpha}-h_{ba}^\alpha h_{ab}^{\mu\alpha}=\epsilon_{ab}^2\left(\mathbf{A}^\alpha_{ab}\partial_\mu\mathbf{A}_{ba}^\alpha-\mathbf{A}^\alpha_{ba}\partial_\mu\mathbf{A}_{ab}^\alpha\right)-\epsilon_{ba}\Delta_{ab}^\alpha\left(\mathbf{A}_{ab}^\alpha\mathbf{A}_{ba}^\mu-\mathbf{A}_{ba}^\alpha\mathbf{A}_{ab}^\mu\right)\\-i\epsilon_{ba}\left(-\epsilon_{ad}\mathbf{A}^\alpha_{ab}\mathbf{A}_{bd}^\mu\mathbf{A}_{da}^\alpha-\epsilon_{bd}\mathbf{A}^\alpha_{ba}\mathbf{A}_{ad}^\mu\mathbf{A}_{db}^\alpha\right)-i\epsilon_{ba}\left(\epsilon_{db}\mathbf{A}^\alpha_{ab}\mathbf{A}^\mu_{da}\mathbf{A}^\alpha_{bd}+\epsilon_{da}\mathbf{A}^\alpha_{ba}\mathbf{A}^\mu_{db}\mathbf{A}^\alpha_{ad}\right)
\end{split}
\end{equation}
It can be written as
\begin{equation}
\begin{split}
h_{ab}^\alpha h_{ba}^{\mu\alpha}-h_{ba}^\alpha h_{ab}^{\mu\alpha}=2i\epsilon_{ab}^2\left(|\mathbf{A}^\alpha_{ab}|^2\partial_\mu\mathbf{\phi}_{ba}^\alpha\right)-\epsilon_{ba}\Delta_{ab}^\alpha\left(\mathbf{A}_{ab}^\alpha\mathbf{A}_{ba}^\mu-\mathbf{A}_{ba}^\alpha\mathbf{A}_{ab}^\mu\right)\\-i\epsilon_{ba}\left(-\epsilon_{ab}\mathbf{A}^\alpha_{ab}\mathbf{A}_{bb}^\mu\mathbf{A}_{ba}^\alpha-\epsilon_{ba}\mathbf{A}^\alpha_{ba}\mathbf{A}_{aa}^\mu\mathbf{A}_{ab}^\alpha\right)-i\epsilon_{ba}\left(\epsilon_{ab}\mathbf{A}^\alpha_{ab}\mathbf{A}^\mu_{aa}\mathbf{A}^\alpha_{ba}+\epsilon_{ba}\mathbf{A}^\alpha_{ba}\mathbf{A}^\mu_{bb}\mathbf{A}^\alpha_{ab}\right)\\\sum_{d\ne a,b}-i\epsilon_{ba}\left(-\epsilon_{ad}\mathbf{A}^\alpha_{ab}\mathbf{A}_{bd}^\mu\mathbf{A}_{da}^\alpha-\epsilon_{bd}\mathbf{A}^\alpha_{ba}\mathbf{A}_{ad}^\mu\mathbf{A}_{db}^\alpha\right)-i\epsilon_{ba}\left(\epsilon_{db}\mathbf{A}^\alpha_{ab}\mathbf{A}^\mu_{da}\mathbf{A}^\alpha_{bd}+\epsilon_{da}\mathbf{A}^\alpha_{ba}\mathbf{A}^\mu_{db}\mathbf{A}^\alpha_{ad}\right)
\end{split}
\end{equation}
where $\phi_{ba}^\mu=\text{Arg}[\mathbf{A}_{ba}^\mu]$, and simplifying it further we get
\begin{equation}
\begin{split}
h_{ab}^\alpha h_{ba}^{\mu\alpha}-h_{ba}^\alpha h_{ab}^{\mu\alpha}=2i\epsilon_{ab}^2\left(|\mathbf{A}^\alpha_{ab}|^2\partial_\mu\mathbf{\phi}_{ba}^\alpha\right)-2i\epsilon_{ba}^2\left|\mathbf{A}^\alpha_{ab}\right|^2\left(\mathbf{A}_{bb}^\mu-\mathbf{A}_{aa}^\mu\right)-\epsilon_{ba}\Delta_{ab}^\alpha\left(\mathbf{A}_{ab}^\alpha\mathbf{A}_{ba}^\mu-\mathbf{A}_{ba}^\alpha\mathbf{A}_{ab}^\mu\right)\\\sum_{d\ne a,b}-i\epsilon_{ba}\left(-\epsilon_{ad}\mathbf{A}^\alpha_{ab}\mathbf{A}_{bd}^\mu\mathbf{A}_{da}^\alpha-\epsilon_{bd}\mathbf{A}^\alpha_{ba}\mathbf{A}_{ad}^\mu\mathbf{A}_{db}^\alpha\right)-i\epsilon_{ba}\left(\epsilon_{db}\mathbf{A}^\alpha_{ab}\mathbf{A}^\mu_{da}\mathbf{A}^\alpha_{bd}+\epsilon_{da}\mathbf{A}^\alpha_{ba}\mathbf{A}^\mu_{db}\mathbf{A}^\alpha_{ad}\right)
\end{split}
\end{equation}
Now, we can further simplify it by using $h_{mn}^\gamma=i\epsilon_{nm}\mathbf{A}_{mn}^\gamma$ for $m\ne n$, 
\begin{equation}
\begin{split}
h_{ab}^\alpha h_{ba}^{\mu\alpha}-h_{ba}^\alpha h_{ab}^{\mu\alpha}=-2i\epsilon_{ab}^2|\mathbf{A}^\alpha_{ab}|^2\underbrace{\left(\mathbf{A}_{bb}^\mu-\mathbf{A}_{aa}^\mu-\partial_\mu\mathbf{\phi}_{ba}^\alpha\right)}_{\text{Shift vector, } \mathbf{S}_{ba}^\mu}-\epsilon_{ba}\Delta_{ab}^\alpha\left(\mathbf{A}_{ab}^\alpha\mathbf{A}_{ba}^\mu-\mathbf{A}_{ba}^\alpha\mathbf{A}_{ab}^\mu\right)\\\sum_{d\ne a,b}\left(-\epsilon_{ad}h^\alpha_{ab}\mathbf{A}_{bd}^\mu \mathbf{A}_{da}^\alpha+\epsilon_{bd}h^\alpha_{ba}\mathbf{A}_{ad}^\mu \mathbf{A}_{db}^\alpha\right)+\left(\epsilon_{db}h^\alpha_{ab}\mathbf{A}^\mu_{da}\mathbf{A}^\alpha_{bd}-\epsilon_{da}h^\alpha_{ba}\mathbf{A}^\mu_{db}\mathbf{A}^\alpha_{ad}\right).
\end{split}
\end{equation}
It can be simplified further
\begin{equation}
\begin{split}
h_{ab}^\alpha h_{ba}^{\mu\alpha}-h_{ba}^\alpha h_{ab}^{\mu\alpha}=-2i\epsilon_{ab}^2|\mathbf{A}^\alpha_{ab}|^2\mathbf{S}_{ba}^\mu+\Delta_{ab}^\alpha\left(\frac{1}{\epsilon_{ab}}h_{ab}^\alpha h_{ba}^\mu+\frac{1}{\epsilon_{ba}}h_{ba}^\alpha h_{ab}^\mu\right)\\\sum_{d\ne a,b}\left(-\frac{1}{\epsilon_{bd}}h^\alpha_{ab}h_{bd}^\mu h_{da}^\alpha+\frac{1}{\epsilon_{ad}}h^\alpha_{ba}h_{ad}^\mu h_{db}^\alpha\right)+\left(\frac{1}{\epsilon_{da}}h^\alpha_{ab}h^\mu_{da}h^\alpha_{bd}-\frac{1}{\epsilon_{db}}h^\alpha_{ba}h^\mu_{db}h^\alpha_{ad}\right).
\end{split}
\end{equation}
Now substituting it back in $\delta(\omega-\epsilon_{ab})$ part of Eq.~\ref{sndthrd}, we get the contribution of 2nd and 3rd term of Eq.~\ref{shiftcurrent1}
\begin{equation}
\begin{split}
\sigma^{23}_{\delta(\omega-\epsilon_{ab})}=\frac{2\pi e^3}{\hbar^2}\int [d\mathbf{k}]f_{ab}|\mathbf{A}^\alpha_{ab}|^2\mathbf{S}_{ba}^\mu\delta(\omega-\epsilon_{ab})+\boxed{\frac{2\pi e^3}{\hbar^2\omega^2}\int [d\mathbf{k}]f_{ab}\Delta_{ab}^\alpha\left(\frac{1}{\epsilon_{ab}}h_{ab}^\alpha h_{ba}^\mu+\frac{1}{\epsilon_{ba}}h_{ba}^\alpha h_{ab}^\mu\right)i\delta(\omega-\epsilon_{ab})}\\\boxed{+\frac{2\pi e^3}{\hbar^2\omega^2}\sum_{d\ne a,b}\int [d\mathbf{k}]f_{ab}\left[\left(-\frac{1}{\epsilon_{bd}}h^\alpha_{ab}h_{bd}^\mu h_{da}^\alpha+\frac{1}{\epsilon_{ad}}h^\alpha_{ba}h_{ad}^\mu h_{db}^\alpha\right)+\left(\frac{1}{\epsilon_{da}}h^\alpha_{ab}h^\mu_{da}h^\alpha_{bd}-\frac{1}{\epsilon_{db}}h^\alpha_{ba}h^\mu_{db}h^\alpha_{ad}\right)\right]i\delta(\omega-\epsilon_{ab})}
\label{s23box}
\end{split}
\end{equation}
It is worth mentioning that the quantity $ \Delta_{ab}^\alpha\left(\frac{1}{\epsilon_{ab}}h_{ab}^\alpha h_{ba}^\mu+\frac{1}{\epsilon_{ba}}h_{ba}^\alpha h_{ab}^\mu\right)$ and\\ $\left[\left(-\frac{1}{\epsilon_{bd}}h^\alpha_{ab}h_{bd}^\mu h_{da}^\alpha+\frac{1}{\epsilon_{ad}}h^\alpha_{ba}h_{ad}^\mu h_{db}^\alpha\right)+\left(\frac{1}{\epsilon_{da}}h^\alpha_{ab}h^\mu_{da}h^\alpha_{bd}-\frac{1}{\epsilon_{db}}h^\alpha_{ba}h^\mu_{db}h^\alpha_{ad}\right)\right]$ are imaginary by default.  This shows that the second and third term of  Eq.~\ref{shiftcurrent1} contains not only the shift vector term but also a few extra terms which include three velocity elements. Next, we would like to check if these extra terms shown in the box above cancel out $\sigma^{56}$ (5th and 6th terms) of Eq.~\ref{shiftcurrent1}. We have
\begin{equation}
\sigma^{56}=\frac{2\pi e^3}{\hbar^2\omega^2}\int [d\mathbf{k}]\frac{h_{ab}^{\alpha}h_{bc}^\alpha h_{ca}^\mu}{\epsilon_{ac}}\left[\frac{f_{ab}}{\omega+i\eta-\epsilon_{ba}}+\frac{f_{cb}}{-\omega+i\eta-\epsilon_{cb}}\right]+\frac{h_{ab}^{\alpha}h_{bc}^\alpha h_{ca}^\mu}{\epsilon_{ac}}\left[\frac{f_{ab}}{-\omega+i\eta-\epsilon_{ba}}+\frac{f_{cb}}{\omega+i\eta-\epsilon_{cb}}\right]
\end{equation}
and after switching $a\leftrightarrow c$ in 3rd and 4th term,it can be written as 
\begin{equation}
\sigma^{56}=\frac{2\pi e^3}{\hbar^2\omega^2}\int [d\mathbf{k}]\frac{f_{ab}}{\epsilon_{ac}}\frac{h_{ab}^{\alpha}h_{bc}^\alpha h_{ca}^\mu}{\omega+i\eta-\epsilon_{ba}}+\frac{f_{ab}}{\epsilon_{ca}}\frac{h_{cb}^{\alpha}h_{ba}^\alpha h_{ac}^\mu}{-\omega+i\eta-\epsilon_{ab}}+\frac{f_{ab}}{\epsilon_{ac}}\frac{h_{ab}^{\alpha}h_{bc}^\alpha h_{ca}^\mu}{-\omega+i\eta-\epsilon_{ba}}+\frac{f_{ab}}{\epsilon_{ca}}\frac{h_{cb}^{\alpha}h_{ba}^\alpha h_{ac}^\mu}{\omega+i\eta-\epsilon_{ab}}
\end{equation}

\begin{equation}
\begin{split}
\implies \sigma^{56}=\frac{2\pi e^3}{\hbar^2\omega^2}\int [d\mathbf{k}]f_{ba}\left[\frac{h_{ba}^{\alpha}h_{ac}^\alpha h_{cb}^\mu}{\epsilon_{bc}}-\frac{h_{ca}^{\alpha}h_{ab}^\alpha h_{bc}^\mu}{\epsilon_{cb}}+\frac{h_{ab}^{\alpha}h_{bc}^\alpha h_{ca}^\mu}{\epsilon_{ac}}-\frac{h_{cb}^{\alpha}h_{ba}^\alpha h_{ac}^\mu}{\epsilon_{ca}}\right]P\left(\frac{1}{\omega-\epsilon_{ab}}\right)\\+\frac{2\pi e^3}{\hbar^2\omega^2}\int [d\mathbf{k}]f_{ba}\left[\frac{h_{ba}^{\alpha}h_{ac}^\alpha h_{cb}^\mu}{\epsilon_{bc}}+\frac{h_{ca}^{\alpha}h_{ab}^\alpha h_{bc}^\mu}{\epsilon_{cb}}-\frac{h_{ab}^{\alpha}h_{bc}^\alpha h_{ca}^\mu}{\epsilon_{ac}}-\frac{h_{cb}^{\alpha}h_{ba}^\alpha h_{ac}^\mu}{\epsilon_{ca}}\right]i\pi\delta(\omega-\epsilon_{ab}).
\end{split}
\end{equation}
Now, the term involving $\delta(\omega-\epsilon_{ab})$ can be written as
\begin{equation}
\sigma^{56}_{\delta(\omega-\epsilon_{ab})}=\frac{2\pi e^3}{\hbar^2\omega^2}\int [d\mathbf{k}]f_{ba}\left[\frac{h_{ba}^{\alpha}h_{ad}^\alpha h_{db}^\mu}{\epsilon_{bd}}+\frac{h_{da}^{\alpha}h_{ab}^\alpha h_{bd}^\mu}{\epsilon_{db}}-\frac{h_{ab}^{\alpha}h_{bd}^\alpha h_{da}^\mu}{\epsilon_{ad}}-\frac{h_{db}^{\alpha}h_{ba}^\alpha h_{ad}^\mu}{\epsilon_{da}}\right]i\pi\delta(\omega-\epsilon_{ab})
\end{equation}
After rearranging these terms and using $\Delta_{ab}^\alpha=h_{aa}^\alpha-h_{bb}^\alpha$, we get
\begin{equation}
\begin{split}
\sigma^{56}_{\delta(\omega-\epsilon_{ab})}=\boxed{(-1)\frac{2\pi e^3}{\hbar^2\omega^2}\int [d\mathbf{k}]f_{ab}\Delta_{ab}^{\alpha}\left[\frac{h_{ab}^\alpha h_{ba}^\mu}{\epsilon_{ab}}+\frac{h_{ba}^{\alpha} h_{ab}^\mu}{\epsilon_{ba}}\right]i\pi\delta(\omega-\epsilon_{ab})}\\+\boxed{(-1)\frac{2\pi e^3}{\hbar^2\omega^2}\int [d\mathbf{k}]f_{ab}\sum_{d\ne a, b}\left[-\frac{h_{da}^{\alpha}h_{ab}^\alpha h_{bd}^\mu}{\epsilon_{bd}}+\frac{h_{db}^{\alpha}h_{ba}^\alpha h_{ad}^\mu}{\epsilon_{ad}}+\frac{h_{ab}^{\alpha}h_{bd}^\alpha h_{da}^\mu}{\epsilon_{da}}-\frac{h_{ba}^{\alpha}h_{ad}^\alpha h_{db}^\mu}{\epsilon_{db}}\right]i\pi\delta(\omega-\epsilon_{ab})}.
\end{split}
\end{equation}
Now, we can see that the above expression $\sigma^{56}_{\delta(\omega-\epsilon_{ab})}$ is equal and opposite to the boxed part (three velocity terms) of Eq.~\ref{s23box}. In other words:
\begin{equation}
\sigma^{23}_{\delta(\omega-\epsilon_{ab})}+\sigma^{56}_{\delta(\omega-\epsilon_{ab})}=\frac{2\pi e^3}{\hbar^2}\int [d\mathbf{k}]f_{ab}|\mathbf{A}^\alpha_{ab}|^2\mathbf{S}_{ba}^{\mu\alpha}\delta(\omega-\epsilon_{ab})
\end{equation}
which is the shift-current expression used in the main text.

\end{widetext}
\end{document}